\documentstyle{l-aa}

\begin{document}
\thesaurus{11(03.13.2; 11.03.1; 11.03.4 A1677, 1305.4+2941)}

\title{An objective and automatic cluster finder:\\
an improvement of the matched-filter method}
\author{W. Kawasaki\inst{1}
   \and K. Shimasaku\inst{1,2}
   \and M. Doi\inst{1,2}
   \and S. Okamura\inst{1,2}}
\institute{Department of Astronomy, the University of Tokyo,
           7-3-1, Hongo, Bunkyo-ku, Tokyo, 113-0033, Japan
   \and    Research Center for the Early Universe, the University of Tokyo,
           7-3-1, Hongo, Bunkyo-ku, Tokyo, 113-0033, Japan}
\offprints{W. Kawasaki (kawasaki@astron.s.u-tokyo.ac.jp)}
\date{Received May 23, 1997; accepted January 30, 1998}

\maketitle

\begin{abstract}

We describe an objective and automated method for detecting clusters of galaxies from optical imaging data. 
This method is a variant of the so-called `matched-filter' technique pioneered by Postman et al. (1996). 
Simultaneously using positions and apparent magnitudes of galaxies, this method can, not only find cluster candidates, but also estimate their redshifts and richnesses as byproducts of detection. 
We examine errors in the estimation of cluster's position, redshift, and richness with a number of Monte Carlo simulations. 
No systematic discrepancies between the true and estimated values are seen for either redshift or richness. 
For clusters with $z$=0.2 and with richness similar to that of the Coma cluster, typical errors in the estimation of position, redshift, and richness are evaluated as $\Delta\theta\sim 10''$ (one third of the projected core radius), $\Delta z\sim$0.02, and $\Delta N/N\sim$12\%, respectively. 
Spurious detection rate of the method is about less than 10\% of those of conventional ones which use only surface density of galaxies. 
A cluster survey in the North Galactic Pole region is executed to verify the performance characteristics of the method with real data. 
Despite poor quality of the data, two known real clusters are successfully detected. 
No unknown cluster with low or medium redshift ($z\leq$0.3) is detected. 
We expect these methods based on `matched-filter' technique to be essential tools for compiling large and homogeneous optically-selected cluster catalogs. 


\end{abstract}

\section{Introduction}

Studies on clusters of galaxies provide us with valuable information on cosmology and extragalactic astronomy. 
It is a common way in most of the studies to collect samples from available catalogs. 
For example, Bahcall (1988) compiled the previous work on two-point angular cluster-cluster correlation function based on published cluster catalogs. 
Rhoads, Gott and Postman (1994) measured a genus curve of Abell clusters for topological studies on the large-scale structure of the Universe. 
Struble and Ftaclas (1994) studied correlations amongst richness, flattening, and velocity dispersion of 350 Abell clusters. 
A large number of reports have also been made on the relations between various properties of clusters (Henry and Tucker 1979; Edge and Stewart 1991; Lubin and Bahcall 1993; Annis 1996; and references therein). 
Multicolor photometry reveals the color evolution of individual galaxies in clusters: Butcher and Oemler (1978, 1984) reported an increasing fraction of `blue' galaxies in clusters with redshift. 
This is known as `Butcher-Oemler effect' and thought to be some sign of galaxy evolution (see also Rakos and Schombert 1995). 
Anyhow, it is indispensable to use large and statistically complete catalogs of clusters for statistical investigations. 

Clusters of galaxies are identified not only as `clusters of galaxies' as it is but also as hot plasma balls. 
Accordingly, both optically- and X-ray-selected cluster catalogs have been constructed so far. 
A number of X-ray clusters were detected by Extended Medium Sensitivity Survey with {\em Einstein} Observatory (Gioia et al. 1990) and {\em ROSAT} All-Sky Survey (Voges et al. 1996). 
X-ray surveys enable one to produce almost complete catalogs of nearby ($z\la$0.2) clusters since it is easy to detect clusters as they are extended X-ray sources. 
At present, however, it is quite difficult to execute a deep X-ray survey over a wide area in the sky to assemble a sufficiently large and complete sample of distant X-ray clusters whereas searching with optical data can reach even more distant clusters. 

Let us outline the development of optical cluster-finding techniques and the optically-selected cluster catalogs themselves in approximately historical sequence. 
Catalogs of nearby ($z\la$0.2) clusters include those compiled by Abell (1958), Zwicky et al. (1961-68), Shectman (1985, hereafter S85), Abell, Corwin and Olowin (1989, hereafter ACO), Lumsden et al. (1992, hereafter L92), and Dalton et al. (1994, hereafter D94). 
For more distant (0.2$<z<$0.9) ones, there exist four catalogs; Gunn, Hoessel and Oke (1986), Couch et al. (1991), Postman et al. (1996, hereafter P96), and Lidman and Peterson (1996, hereafter LP96). 
All the above catalogs except for S85, L92, D94, P96, and LP96 were constructed by eye selection of clusters on photographic plates. 
We can easily imagine that large efforts were required to assemble these catalogs. 
However, these catalogs are claimed to suffer from inhomogeneity and contamination: a significant fraction of clusters may be missed (for Abell/ACO catalog, see Gunn, Hoessel and Oke 1986; Sutherland 1988; Ebeling et al. 1993) while some of the cataloged clusters may be spurious (Lucey 1983). 
These effects become much more critical for fainter (namely, more distant and/or poorer) ones. 

S85, L92, and D94 detected clusters semi-objectively (L92 and D94 did it also automatically): S85 and L92 employed count-in-cells technique while D94 adopted percolation technique. 
Yet, both techniques use only projected positions of galaxies and simply pick up overdensities in the two-dimensional distribution of galaxies. 
Consequently they cannot quantify the detection rate of spurious clusters due to chance coincidence of galaxies on the sky. 
Collins et al. (1995) and Ebeling and Maddox (1995) reported the significant amounts of contamination in the catalogs compiled by L92 and D94, respectively. 
Furthermore, they pick up overdensities of galaxies within the area of a fixed apparent angular size, despite that the actual angular extension of clusters undoubtedly changes with distance. 
This means that cluster-finding criteria in these methods do change with redshift. 
Thus these catalogs may not be regarded as far more objective than the `classical' ones such as Abell/ACO catalog.

Escalera and MacGillivray (1995, 1996) have searched for structures of various scales, from groups up to superclusters, using wavelet transform. 
Wavelet transform does not stick to a certain apparent size of structures and enables one to execute a `multi-scale' analysis. 
However, using only galaxy positions on the sky, wavelet transform also cannot quantify spurious detection rate. 
Dividing the total sample into subsamples with small ranges of magnitude, as Escalera and MacGillivray (1996) did, may somewhat suppress spurious detections in such methods as count-in-cells, percolation, and wavelet transform, but at the same time, it may also reduce real signal. 

P96 developed and employed an innovative cluster-finding method based on `matched-filter' technique. 
The point of their method is to use both projected positions and apparent magnitudes of galaxies simultaneously. 
This enables one to obtain rough estimates of redshifts and richnesses for detected clusters without any spectroscopic information. 
We need only one broad band images, while obtaining photometric redshift requires more than three bands. 
The catalog by P96 contains 79 distant clusters ($0.2<z<1.2$) from $V$ and $I$ band data over 5 deg$^2$ obtained with 4-Shooter CCD camera (Gunn et al. 1987) attached to Palomar 5m Hale telescope. 
It is noted here that all the above cluster catalogs except for the one by P96 were based on photographic plates. 
Although CCDs appeared as new optical detectors taking the place of photographic plates in 1980s, it was extremely time consuming to make use of them for survey observations because of their small sizes. 
However, recent developments of large-format CCDs and of CCD mosaic cameras made it possible to quickly survey over a wide area ($\sim$some deg$^2$) on the sky and to obtain large amount of data of good quality. 
The cluster catalog compiled by P96 is also the first CCD-based cluster catalog. 
Using similar methods, LP96 conducted a search for distant clusters and built a catalog of 105 candidates from $I$ band CCD data covering 13 deg$^2$, obtained with Anglo-Australian Telescope. 

Prompted by the work of P96, we developed a variant method for automatic and objective cluster-finding with optical imaging data. 
Our method has some nontrivial differences from the one by P96 in the details of detection process, such as binning the input data and employing Poisson statistics. 
These differences have made apparent improvements in processing time and in accuracies in estimating redshift. 
In particular, the systematic discrepancy between true and estimated values found in P96 has been largely reduced. 

In Sect. 2, we discuss the principle of the cluster-finding method. 
Detailed performance tests of the method are described in Sect. 3. 
In Sect. 4, as a performance verification test with real data, we perform a cluster survey using the $B$ band galaxy samples within the 4.9 deg$^2$ region around the North Galactic Pole (NGP), obtained with our Mosaic CCD Camera attached to 1.05m Schmidt Telescope at Kiso Observatory, Japan. 

Throughout this paper, we assume $H_0$ = 80 \,km \,s$^{-1}$ \,Mpc$^{-1}$ and $q_0=0.5$. 

\section{The method}

We detect clusters with maximum likelihood method in a way similar to that by P96, using models of surface density and apparent magnitude distribution of cluster galaxies and field galaxies. 
The cluster model has two free parameters, namely, its redshift and richness. 

\subsection{Models}
\subsubsection{Cluster}

We assume spherical symmetry for simplifying the cluster model. 
The radial distribution of the galaxies in the model cluster matches the King model with $c \equiv \log (r_{\rm tidal}/r_{\rm core}) = 2.25$ (King 1966; Ichikawa 1986) and $r_{\rm core} = 170h^{-1}{\rm kpc}$ (Girardi et al. 1995), where $r_{\rm core}$ and $r_{\rm tidal}$ are core radius and tidal radius, respectively. 
The luminosity function of the galaxies obeys the Schechter function (Schechter 1976) with $\alpha = -1.25$ and $M_B^* = -20.4$ (converted from $M_{B_J}^* = -20.12$ by Colless 1989, rescaling $H_0$ and using $M_{B_J} = M_B - 0.18$ by Yoshii and Takahara 1988). 
Morphological type mixture, luminosity segregation, substructure, and influences by cD galaxies are not considered. 
To compute apparent features of the model cluster, two more parameters, namely redshift $z_{\rm fil}$ (hereafter we call it `filter redshift' after the manner of P96) and richness $N$, must be assigned. 
Obviously, if $z_{\rm fil}$ is larger, angular extension of the model clusters becomes smaller and member galaxies become fainter. 
Dimming by the $K$-correction effect is assumed to be $K_B(z_{\rm fil}) = 4.7 \times z_{\rm fil}$ (for $z_{\rm fil}\la$0.6), which is the value for elliptical galaxies (Fukugita, Shimasaku and Ichikawa 1995). 
We define $N$ to be the number of all galaxies brighter than ($M^*+5$). 
The parameter $N$ roughly represents the population of bright, giant galaxies in a cluster, ignoring dwarf galaxies whose natures such as spatial distribution or luminosity function are still unclear. 

\subsubsection{Field}

We assume that field galaxies are randomly distributed on the sky, namely, angular two-point correlation function is not considered. 
For simulations in Sect. 3, we adopt deep galaxy number count data by Metcalfe et al. (1995) as the model of apparent magnitude distribution of field (foreground and background) galaxies. 
For the actual galaxy data discussed in Sect. 4, we use the magnitude distribution of all galaxies in the survey area to search clusters as if it were that of pure field galaxies. 
This causes an overestimate of the number of field galaxies, especially when the survey area is small and there is a cluster covering a bulk portion of the area by chance. 
Hence we have to deal with an enough large area so that clusters or even a large-scale structure will not seriously affect the estimate of the number of field galaxies in the area. 
Yet sometimes iteration would be needed. 
For that case, we mark conspicuous `cluster' regions by referring to the first-time result and then execute the second calculation using the `more accurate' field galaxy sample in the area except for the `cluster' regions. 

\subsection{Algorithm}

What we need at the beginning is just a usual catalog of galaxies containing projected positions and apparent magnitudes. 
Fig. 1 shows an example galaxy distribution. 
The galaxies are generated by a Monte Carlo simulation based on the model described in Sect. 2.1. 
A cluster with $(z,N)=(0.20,1000)$, roughly equal to an Abell richness class 0-1 cluster, is located at the center. 

Next we compute the likelihood ${\cal L}$ that a cluster is present at a particular point. 
We consider $n_{\theta}$ concentric annular regions centered on the position. 
Their angular inner radii and widths are $\theta _i$ and $\Delta \theta _i (1\leq i\leq n_{\theta})$, respectively. 
By counting the number of galaxies which fall in $n_m$ magnitude bins ($m_j\leq m<m_j+\Delta m_j\; (1\leq j\leq n_m)$) for each annular region, we obtain an array $O_{ij}\; (1\leq i\leq n_{\theta}, 1\leq j\leq n_m)$ consisting of $n_{\theta} \times n_m$ galaxy numbers. 

On the other hand, we can calculate an equivalent array $M_{ij}$ for the model galaxies. 
$M_{ij}$ is described as 
\begin{equation}
M_{ij} = N\; C_{ij} + F_{ij},
\end{equation}
where $C_{ij}$ is an array for member galaxies in a normalized model cluster located at the center of $n_{\theta}$ concentric annular regions, and $F_{ij}$ is an array for field galaxies. 
$C_{ij}$ is written as 
\begin{equation}
C_{ij} = 2\pi \int _{\theta _i}^{\theta _i+\Delta \theta _i} \theta \sigma _c(\theta )\; d\theta \; \int _{m_j}^{m_j+\Delta m_j} \phi _c(m)\; dm,
\end{equation}
where $\sigma _c(\theta )$ is surface density profile and $\phi _c(m)$ is differential luminosity function of cluster galaxies. 
Both $\sigma _c(\theta )$ and $\phi _c(m)$ depend on filter redshift and are normalized as 
\begin{equation}
2\pi \int _0^{\theta _{\rm tidal}} \theta \sigma _c(\theta )\; d\theta = 1
\end{equation}
and 
\begin{equation}
\int _{-\infty }^{m^*+5} \phi _c(m)\; dm = 1.
\end{equation}
$F_{ij}$ is written as 
\begin{equation}
F_{ij} = 2\pi \sigma _f \int _{\theta _i}^{\theta _i+\Delta \theta _i} \theta \; d\theta \; \int _{m_j}^{m_j+\Delta m_j} \phi _f(m)\; dm,
\end{equation}
where $\sigma _f$ is surface density and $\phi _f(m)$ is differential luminosity distribution of field galaxies. 

The logarithmic likelihood is given by 
\begin{displaymath}
\ln {\cal L} = \sum _{i, j} \ln \left\{ \frac{M_{ij}^{O_{ij}}\; e^{-M_{ij}}}{O_{ij}!} \right\}
\end{displaymath}
\begin{equation}
= \sum _{i, j} \left\{ O_{ij}\ln (N\; C_{ij}+F_{ij})-(N\; C_{ij}+F_{ij})-\ln (O_{ij}!) \right\}.
\label{likelihood}
\end{equation}

Here we assume that $O_{ij}$ obeys Poisson statistics since their values amount to about 10 or less for typical clusters at $z\la$0.2 with our choice of values for $\Delta \theta _i$ and  $\Delta m_j$ (see the next section). 
If we assume Gaussian distribution for $O_{ij}$, Eq. (\ref{likelihood}) becomes simply equivalent to $-\chi ^2$ (P96 did so, expecting that there would be enough background galaxies. See Eq. 12 of their paper).
However, this assumption leads us to overestimating $N$ about 20\% of the true value for the case of ($z,N$)=(0.20,1000). 
This is because Poissonian distribution is not symmetrical and has a longer tail toward larger value. 
The use of Poisson statistics helps to reduce the possible systematic error in the estimation of the redshift and the richness. 
\begin{figure}
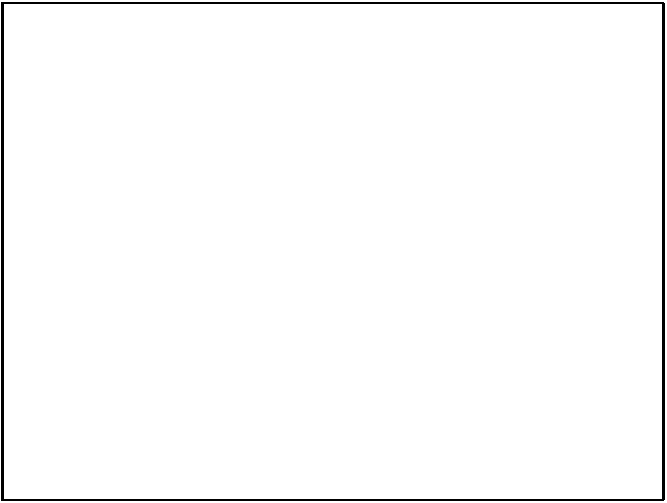

\picplace{6.6cm}
\caption[ ]{Galaxy distribution in an area containing an artificial cluster with ($z, N$) = (0.20, 1000) at the center. The symbol size changes with the apparent magnitude. The largest and smallest symbols correspond to $m_B$ = 16.0 and 23.5, respectively.}
\end{figure}
\begin{figure}
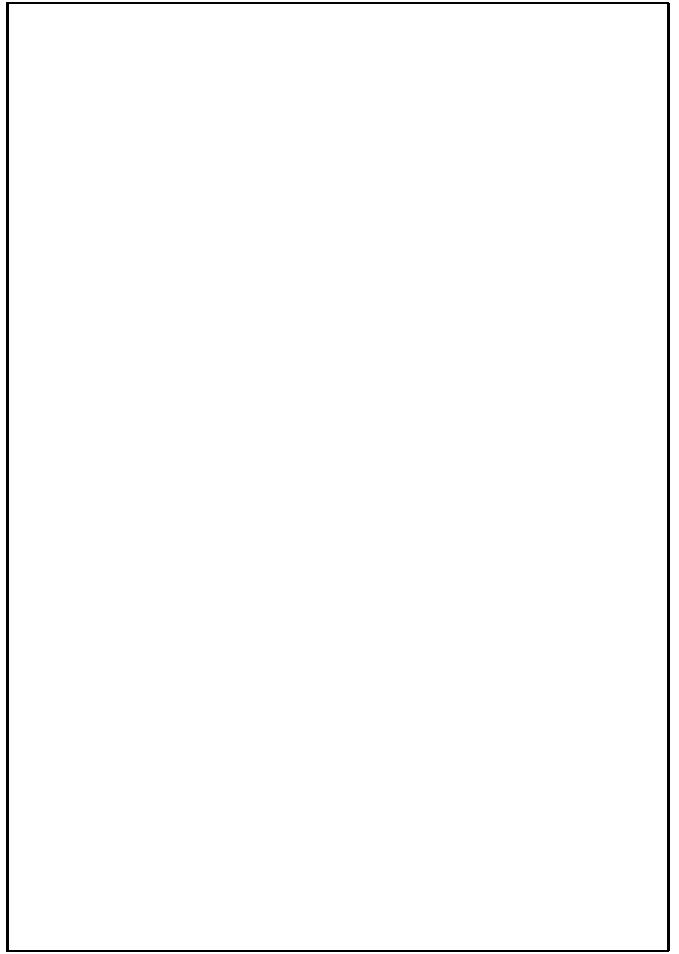

\picplace{12.6cm}
\caption[ ]{`Likelihood image' (upper panel ({\bf a})) and `Richness image' (lower panel ({\bf b})) of the artificial cluster in Fig. 1 for $z_{\rm fil}$ = 0.20.}
\end{figure}

Eq. (\ref{likelihood}) is a function of both filter redshift $z_{\rm fil}$ and richness $N$. 
In order to simplify calculations, we first fix $z_{\rm fil}$ to a certain value and maximize ${\cal L}$ by optimizing only $N$. 
The partial derivative of the logarithmic likelihood (\ref{likelihood}) with respect to $N$ is 
\begin{equation}
\frac{\partial }{\partial N}\ln {\cal L} = \sum _{i, j} \left\{ \frac{O_{ij}C_{ij}}{N\; C_{ij}+F_{ij}} - C_{ij} \right\}.
\label{dif_likelihood}
\end{equation}
Eq. (\ref{dif_likelihood}) is apparently a monotonically decreasing function of $N$. 
If we find a certain richness value $N_{\rm p}$ for which Eq. (\ref{dif_likelihood}) becomes zero, ${\cal L}$ has a peak value ${\cal L}_{\rm p}$ at $N=N_{\rm p}$. 
Computing ${\cal L}_{\rm p}$ and $N_{\rm p}$ at every point in the whole image, we obtain a `likelihood image' ${\cal L}_{\rm p} (x,y)$ and a `richness image' $N_{\rm p} (x,y)$ for the fixed filter redshift. 

Figs. 2a and 2b, respectively, show the `likelihood image' and the `richness image' for $z_{\rm fil}=0.20$ generated from the galaxy distribution shown in Fig. 1. 
We can recognize the existence of a cluster by a peak in both images. 
However, appearances of the peaks are quite different. 
While the peak in the `richness image' is simple and very prominent, there exists a ring-like region of slightly lower likelihood around a weak peak in the center of the `likelihood image', and ${\cal L}_{\rm p}$ increases again toward further out of the ring. 
This is because it is difficult to discriminate a cluster with very small $N$ from `field'. 
Though only `likelihood image' is theoretically needed to detect clusters, peaks in `likelihood image' corresponding to clusters are often very obscure as seen in Fig. 2a. 
We therefore find a peak in `richness image' at first, and then check if there is also a peak in corresponding `likelihood image'. 
If a peak exists at nearly the same point in both images, we regard it as a cluster candidate. 

We obtain several pairs of `likelihood image' and `richness image' and find peaks in both images for other filter redshifts. 
Then we plot the peak ${\cal L}_{\rm p}$ and the peak $N_{\rm p}$ as functions of filter redshift for each cluster candidate. 
An example is shown in Figs. 3a and 3b, respectively. 
If we find a peak in ${\cal L}_{\rm p}-z_{\rm fil}$ plot (Fig. 3a), $z_{\rm fil}$ at the peak is the redshift estimate of the cluster candidate (hereafter $z_{\rm est}$). 
Once $z_{\rm est}$ is obtained, $N_{\rm p}$ for that $z_{\rm est}$ (hereafter $N_{\rm est}$) can also be found as shown in Fig. 3b. 
Some cluster candidates do not show any remarkable single peak of ${\cal L}_{\rm p}$ in the ${\cal L}_{\rm p}-z_{\rm fil}$ plot. 
Such candidates may be spurious. 

\section{Performance test}

We examine the performance of our method described in Sect. 2 by Monte Carlo simulations. 
The errors in estimates of position, redshift, and richness, missing rate of existing clusters (incompleteness), and spurious detection rate are investigated. 
Some comparison of this method with that by P96 is also discussed. 
In this section, we adopt $\theta _1 = 0$, $\Delta \theta = 2r_{\rm core}/d_A(z=0.15)$ where $d_A(z)$ is angular diameter distance, $n_{\theta}$ = 5, $m_1$ (in the $B$ band) = 14.0, $\Delta m$ = 0.5, and $n_m$ = 19. 
Limiting magnitude is set to $m_B$=23.5. 

\subsection{Estimates of position, redshift, and richness}
\subsubsection{Monte Carlo simulation}

When a cluster is detected, its projected position, redshift, and richness are estimated. 
Errors in these estimates depend not only on the real redshift and richness (hereafter $z_{\rm real}$ and $N_{\rm real}$, respectively), but also on limiting magnitude, color band, and the Galactic absorption. 
To evaluate the dependence on $z_{\rm real}$ and $N_{\rm real}$, we examine 20 cases with $z_{\rm real}$=\{0.16, 0.20, 0.24, 0.28\} for $N$=300 and $z_{\rm real}$=\{0.16, 0.20, 0.24, 0.28, 0.35, 0.40, 0.45, 0.50\} for $N$=\{1000, 3000\}. 
In the present study we limit the redshift range to $z\leq 0.5$, for which we expect that ample data will be available in the near future. 
For each case, 500 artificial $B$ band galaxy samples are generated by Monte Carlo simulation according to the model described in Sect. 2. 
$N$=300 corresponds to MKW-AWM systems (Bahcall 1980), $N$=1000 corresponds to Abell richness class 0-1, and $N$=3000 corresponds to Abell richness class 2 (similar to the Coma cluster). 
The relationship between our $N$ and Abell richness parameter $c \equiv N_{m_3\leq m\leq m_3+2}$ is presented in Appendix. 
The Galactic absorption is not considered. 

\subsubsection{Position}

We measure angular distance between the true position of the cluster center $\vec{x}_0$ and the estimated position $\vec{x}_N$ where $N_{\rm p}$ is maximum in the `richness image' for $z_{\rm est}$. 
Properly speaking, we must use the position $\vec{x}_{\cal L}$ corresponding to peak ${\cal L}_{\rm p}$, rather than peak $N_{\rm p}$. 
It is, however, much easier to detect a peak in the `richness image' than in the `likelihood image' as described in Sect. 2. Since $\vec{x}_N$ is actually close enough to $\vec{x}_{\cal L}$ (separation is much less than the core radius), there is almost no problem to use $\vec{x}_N$. 
The estimated positions are distributed around $\vec{x}_0$ and are well fit by two-dimensional Gaussian distribution. 
Fig. 4 shows the values of $\sigma_{\rm est}$ of the best-fit Gaussians normalized by the angular core radius. 
The errors in the estimations are about $\theta_{\rm core}$, 0.5 $\theta_{\rm core}$, and 0.3 $\theta_{\rm core}$ for $N$ = 300, 1000, and 3000, respectively. 
These values are quite small compared with the angular extensions of the clusters themselves. 

\begin{figure}
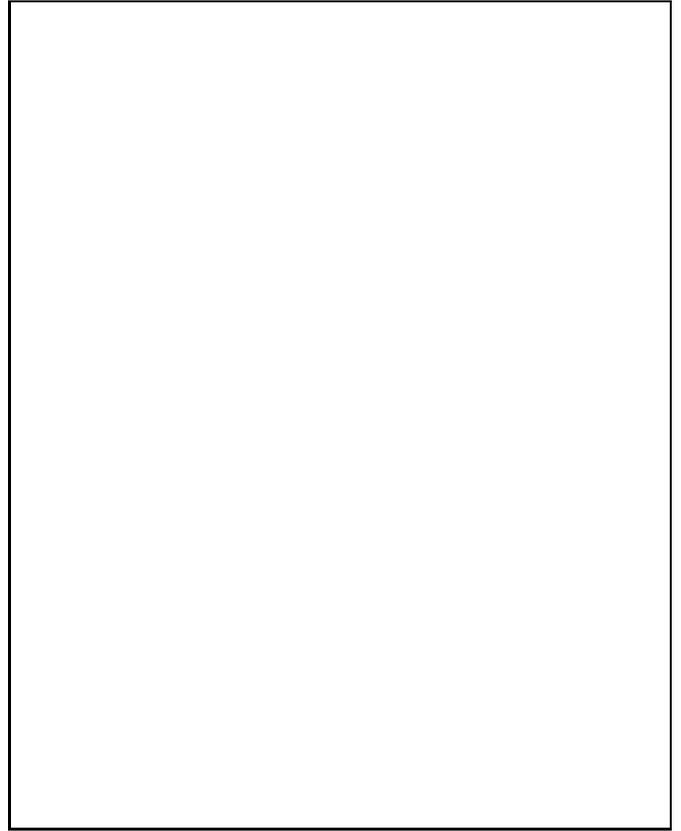

\picplace{11cm}
\caption[ ]{Upper panel ({\bf a}) displays peak logarithmic likelihood as a function of filter redshift for the artificial cluster in Fig. 1. Lower panel ({\bf b}) shows peak richness as a function of filter redshift for the same data.}
\end{figure}
\begin{figure}
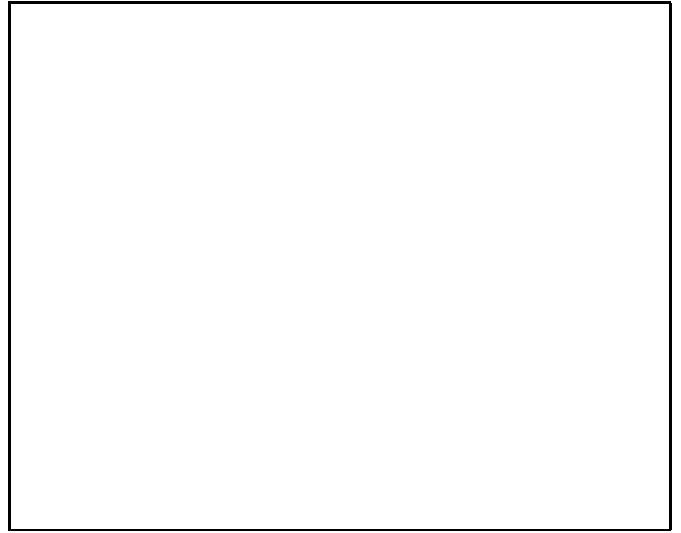

\picplace{7cm}
\caption[ ]{Errors in position estimation $\sigma_{\rm est}$ normalized by the angular core radius $\theta_{\rm core}$ as a function of cluster redshift. Filled circles and solid line are for clusters with $N$ = 3000, open circles and dashed line for $N$ = 1000, and open triangles and dotted line for $N$ = 300. How much fainter we can observe than $m^*$ is shown at the top.}
\end{figure}

\subsubsection{Redshift and richness}

\begin{figure}
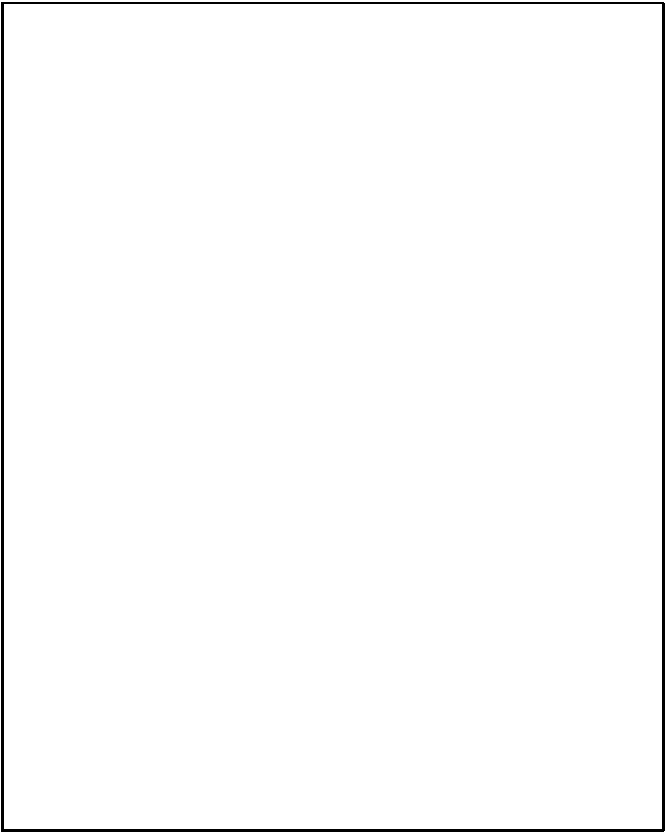

\picplace{11cm}
\caption[ ]{Errors in estimates of redshift and richness. In each panel of a-d, three cases corresponding to $N$ = 3000, 1000 and 300 are shown. The plus mark means the most reliable value. Inner and outer contours around the plus mark show 68\% and 95\% confidence levels, respectively.}
\end{figure}

Fig. 5 shows the result of redshift and richness estimations for the nearby 12 cases of the artificial clusters. 
The plus marks indicate the most probable values and the two contours represent 68\% and 95\% confidence levels. 
Three sets of a plus mark and two contours in each panel are for $N$ = 300, 1000, and 3000. 
The contours are all elongated in the direction from the bottom left to the upper right. 
This is because the estimation of redshift and that of richness are coupled with each other. 
That is, a rich cluster at a large distance looks similar to a less rich, nearer cluster. 

The direction of the largest dispersion in the distributions of 500 points of ($z_{\rm est}$, $N_{\rm est}$), namely, the direction of the major axis of the contours in Fig. 5, differs amongst clusters of different richnesses. 
This is due to different relative ratio of number of cluster galaxies to that of field galaxies within cluster region. 
Figs. 6a and 6b show accuracies in the estimates of redshift and richness, respectively, for all the 20 cases. 
Error bars mean the widths of 68\% confidence contours in Fig. 5, projected onto the corresponding axis. 
Errors in the estimates of redshift and richness at $z$=0.2 are, respectively, about 0.02 and 12\% for $N_{\rm real}$=3000 clusters and about 0.04 and 30\% for $N_{\rm real}$=1000 clusters. 
No systematic deviations from true values are seen. 
Thus, redshift and richness estimations by this method go fairly well without any spectroscopic information. 

\begin{figure}
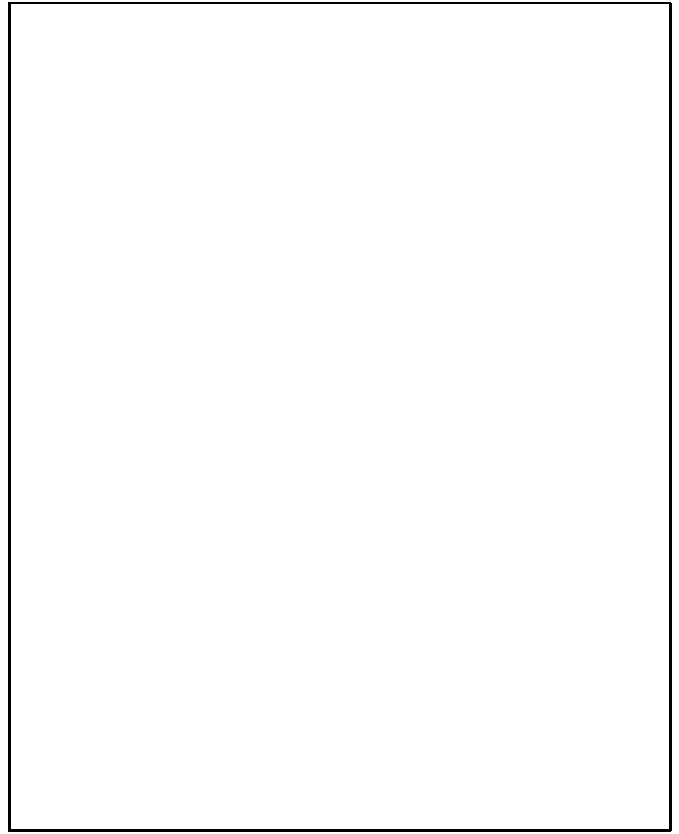

\picplace{11cm}
\caption[ ]{Upper panel ({\bf a}) shows errors in redshift estimation, while lower panel ({\bf b}) shows errors in richness estimation. The error bars represent $\pm 1\sigma$, corresponding to the inner contours in Fig. 5. Filled circles and solid line are for clusters with $N$ = 3000, open circles and dashed line for $N$ = 1000, and open triangles and dotted line for $N$ = 300.}
\end{figure}

These errors are internal. 
In practice, there exist external errors in addition to the internal ones investigated above, owing to intrinsic properties of real clusters: dispersion in $M^*$ values, variations in shapes of luminosity functions and surface density profiles, elongation of clusters, substructures, overlapping with other clusters along the line of sight, etc.. 
These uncertainties will affect the estimations of $z_{\rm est}$ and $N_{\rm est}$. 
Moreover, for very distant ($z\sim 1$) ones, systematic evolutions of cluster galaxies or evolution of clusters themselves may also affect the estimates. 

The most direct and serious effect on the redshift estimation comes from the dispersion in $M^*$. 
Colless (1989) evaluated the upper limit of dispersion in $M^*$ to be 0.4 mag, which corresponds to a redshift estimation error of $\Delta z\sim 0.03$ in $B$ band. 
For other uncertainties, it is difficult to quantitatively evaluate their effects on redshift and richness estimations. 
Intrinsic properties of real clusters are still unclear. 
Therefore, we should rather study them in more detail {\em after} obtaining a `large and statistically complete' cluster catalog by an `objective' cluster-finding method such as the present one by changing parameters of cluster models. 
Spectroscopic observations are also needed to verify the results of redshift estimations and to study $M^*$ values and its dispersion, evolution, etc.. 
Several times of iterations would be needed to establish both a really objective cluster catalog and a really objective cluster-finding technique. 

\subsection{Incompleteness}

For a real but very faint (poor and/or distant) cluster, we may miss either the likelihood peak or the richness peak or both. 
To evaluate probabilities of missing real clusters, we again use the 20$\times$500 artificial clusters. 
We find that our cluster-finding technique can detect almost all clusters up to $z_{\rm real}\sim$ 0.30. 
In the case of $N_{\rm real}$=3000, the missing probabilities do not exceed 0.2\% (namely, no cluster in 500 samples is missed) at $z_{\rm real}\leq$0.35. 
Then the number of missed clusters begins to increase up to $\sim$5\% at $z_{\rm real}$=0.50. 
In the case of $N_{\rm real}$=1000, incompleteness appears at $z_{\rm real}$=0.28 and grows up to $\sim$15\% at $z_{\rm real}$=0.50. 
Even for poor ($N_{\rm real}$=300) clusters, only 8-15\% are missed in the range of 0.16$\leq z_{\rm real}\leq$0.28. 

Gunn, Hoessel and Oke (1986) pointed out the large incompleteness of the Abell catalog at $z\sim$ 0.30. 
There are only 8 Abell clusters in the regions they observed, although they estimated that about 150 clusters exist up to the redshift limit of 0.30. 
Complete sampling of distant clusters is indispensable for the correct understanding of their nature. 

\subsection{Spurious detection rate}

We study the detection rate of non-physical (spurious) clusters using artificial random distribution of galaxies. 
Of course, the actual field galaxies have non-zero angular correlation function (e.g., Davis and Peebles 1983). 
Therefore the actual spurious detection rate may be slightly different from those based on random distribution. 
Even if the distribution of field galaxies is random, certainly there exist some galaxy clumps by projection effects. 
Searching for clusters by simply finding overdensities of galaxies on the sky will result in detecting a number of such spurious ones. 
Here we display how well we can suppress spurious detections by taking into account magnitude information and projected positions simultaneously. 

We evaluate the spurious detection rate with 1000 sets of artificial $50'\times 50'$ `field' data which do not contain any clusters. 
The limiting magnitude in the $B$ band is 23.5. 
In order to evaluate the spurious detection rate rigorously, it is necessary to obtain ($z_{\rm est}, N_{\rm est}$) of all spurious clusters. 
However, since this is a time-consuming task, we adopt a simpler approach to roughly estimate the upper limit of spurious detection rate here. 

In a `richness image' for a given filter redshift, we simply count the number of `richness peaks' which exceed a given threshold value $N_{\rm th}$ and are not separated more than 3$\theta_{\rm core}$ from the corresponding `likelihood peak's. 
We perform this task for the 1000 artificial `fields'. 
The distribution of the 1000 `richness peak's (per $50'\times 50'$ area) is very well fit by Poissonian distribution. 
We compute the best-fit Poissonian mean value ($\lambda$) with least squares method.
Then we convert the $\lambda$ to the value per deg$^2$ and simply regard it as an upper limit of spurious detection rate. 
In Fig. 7, we show the upper limits of spurious detection rate for four thresholds ($N_{\rm th}$=200, 300, 400, and 500) as a function of filter redshift by solid lines. 

To compare these values with those by a traditional method, we calculate the spurious detection rates by count-in-cells technique with cell's size of 2$\theta_{\rm core}$ for 2.5$\sigma$ and 3$\sigma$ levels ($\sigma$ is the standard deviation of the distribution of the number of galaxies per cell). 
They are also shown in Fig. 7 by dashed lines. 
It is clearly seen that the use of magnitude information remarkably suppresses the spurious detection rate, especially at lower redshift. 

Moreover, the values represented by solid lines in Fig. 7 are just upper limits. 
We can further suppress the spurious detection rate by examining the shape of the ${\cal L}_{\rm p}-z_{\rm fil}$ curve. 
For the most of spurious clusters, the ${\cal L}_{\rm p}-z_{\rm fil}$ curves (e.g., Fig. 3a) do not have a single peak and are sometimes very noisy so that we can exclude these cluster candidates as `junks' from the resulting cluster catalog. 
For some of the others, however, the ${\cal L}_{\rm p}-z_{\rm fil}$ curves have a good-looking peak just like the one seen in Fig. 3a. 
These are `really spurious' clusters, which we can not discriminate from real clusters even with additional information of galaxy magnitudes. 

Let us roughly estimate the numbers of `really spurious' clusters with $z$=0.16, 0.20, 0.24, and 0.28. 
For the case of $z$=0.16, first we randomly select 10 spurious cluster candidates in the `richness images' for $z_{\rm fil}$=0.16. 
Then we examine their ${\cal L}_{\rm p}-z_{\rm fil}$ curves to find ones with good-looking peaks, and count the number of `really spurious' clusters, $z_{\rm est}$ of which falls into 0.16$\pm$0.02 (0.02 is half of the interval of $z_{\rm fil}$ for which likelihood values are actually computed). 
The numbers of `really spurious' clusters are found to be 3 and 1 for $z_{\rm fil}$=0.16 and 0.20, respectively. 
No `really spurious' clusters are found for $z_{\rm fil}$=0.24 and 0.28. 
Thereby the correct spurious detection rate goes down to much lower than the upper limit: it is about 30\% at $z$=0.16, 10\% at $z$=0.20, and less than 10\% at $z$=0.24 and 0.28, of the values shown as the solid lines in Fig. 7. 

\begin{figure}
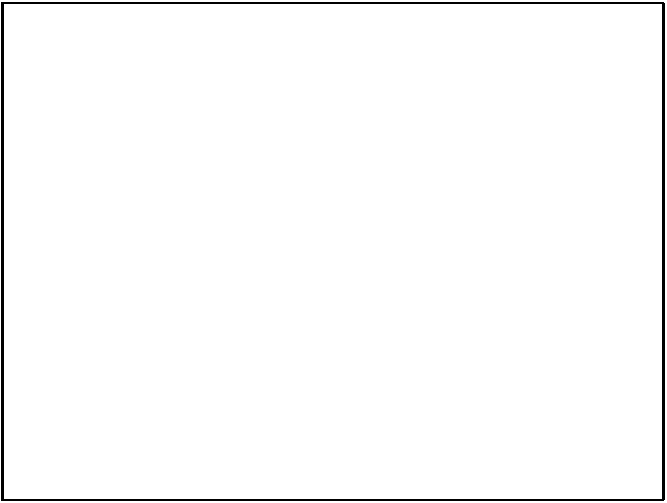

\picplace{6.6cm}
\caption[ ]{Spurious detection rates per deg$^2$ as functions of filter redshift. Solid lines show the results (upper limit) of the present method, while dashed lines show those of count-in-cells technique.}
\end{figure}

Here we examine spurious detection only by simple statistical projection effect. 
In addition to this case, overlaps of two or more poor groups, superpositions of field galaxies on poor groups, and small clumpy portions in outskirts of nearby large clusters (see Sect. 4) also contribute to spurious detection. 
For these cases, ${\cal L}_{\rm p}-z_{\rm fil}$ curves will also be very noisy or have several peaks or no peak. 
Such cluster candidates can easily be excluded from the resulting catalog or checked off as doubtful ones. 
Only spectroscopic observations of the galaxies of these cluster candidates can reveal what in fact they are. 

Even for conspicuous galaxy clumps found by simple glances at galaxy distributions, some of them eventually turn out to be spurious. 
On the other hand, some marginal concentration of galaxies are identified as real clusters. 
Using projected positions and magnitude simultaneously, we often obtain quite different results from those by intuitive methods which use only projected distributions of galaxies. 
In other words, we can quite easily identify a number of non-physical clusters which we can never discriminate without magnitude information. 

\subsection{Comparison with the method by P96}

This method is a variant of the one by P96. 
The basic idea is identical, but there are two main differences in the actual procedures. 

The first one is the form of likelihood function. 
While our likelihood function is based on Poisson statistics (Eq. \ref{likelihood}), the one employed by P96 (Eq. 15 of their paper) is based on Gaussian statistics, namely, P96's likelihood is proportional to $\chi ^2$. 
Of course, the equation is valid, as they say, when there are sufficient number of background galaxies. 
However, especially when the limiting magnitude is brighter and there are fewer galaxies, the adoption of Gaussian statistics becomes unsuitable. 
Moreover, the likelihood function actually employed by P96 (Eq. 16 of their paper) is an approximated shape of the formal expression, though they mention that maximizing the simplified likelihood is roughly equivalent to maximizing the formal one. 
We do not investigate how these differences affect the accuracies of redshift estimation. 
However, the results are obviously different. 
For our case, the true and estimated redshifts agree well and a significant systematic deviation does not appear up to at least $z$=0.5. 
On the other hand, the redshift estimated with the method by P96 tends to be systematically smaller than the true value (see Fig. 14 of their paper). 
Considering the different color band, limiting magnitude, and Hubble constant between this work and P96, $z$=0.5 for our case corresponds to $z\sim$0.7 and $z\sim$1.0 for the cases of $V_4$ and $I_4$ bands, respectively, in their paper. 
At $z_{\rm true}$=1.0 in the lower panel (for $I_4$ band) of the Fig. 14 of P96's paper, the discrepancy between the true redshift and the mean value of the estimated ones are no less than 0.2, while that for our method is much less than 0.01 at $z$=0.50 as shown in Fig. 6a. 

The second difference is binning procedure. 
We bin the galaxies with their positions and magnitudes while P96 did not. 
Binning procedure significantly reduces the processing time (down to a tenth) in the same computational environment. 
This is crucial for constructing an, especially, huge cluster catalog in which such techniques can display their real worth. 

\section{Cluster survey in the NGP region}

In Sect. 3, we have examined the performance of this method with a well-behaved model cluster. 
Further tests with real galaxy data are needed for putting the method to practical use. 
Here we perform a cluster survey with 5.3 deg$^2$ data of the North Galactic Pole region in the $B$ band, which were obtained with our Mosaic CCD Camera 1 (hereafter MCCD1) attached to 1.05m Schmidt telescope at Kiso Observatory, Japan. 

\subsection{Observation}

The observation was made from March 16th to 18th in 1994 at Kiso Observatory. 
MCCD1, consisting of $2\times 8$ TC215 CCDs, was attached to 1.05m Schmidt telescope. 
The CCDs have 1000$\times$1018 pixels and the pixel size is 12$\mu m\times$12$\mu m$. This corresponds to the scale of 0.75 arcsec/pixel at the prime focus of Kiso Schmidt telescope (see Sekiguchi et al. 1992 for more details). 
In MCCD1, CCD chips are placed with large intervals between them. 
Therefore we have to take 15 exposures to obtain data for a contiguous region on the sky. 

The data are centered at ($\alpha$,$\delta$) = (13$^{\rm h}$09$^{\rm m}$.1, +29$\degr$ 48.3$\arcmin$) (J2000.0), covering $1.7\times 3.4$ deg$^2$ with 15 exposures. 
Unfortunately, seeing was poor amounting up to 6.0 arcseconds. 
A chip is out of work and the data lack in the south-eastern corner of about 0.3 deg$^2$, hence the actual observed area is 5.3 deg$^2$. 

\subsection{Galaxy catalog construction}
\subsubsection{Data reduction}

The data reduction is executed in a usual way for optical CCD imaging data. 
After bias subtraction, flat-fielding and sky subtraction, we measure relative positions and relative gains between all pairs of neighboring frames taken either with the same CCD or with the adjacent CCDs at different exposures, using stars common in both frames, to construct a mosaicked image. 
When matching the images, we made positional and flux errors uniformly spread over the whole data. 
Typical seeing size is 3.5-4.0 arcseconds, but among the 15 exposures, there are some data with large seeing ($\sim$ 6.0 arcseconds). 
To keep homogeneity in detecting objects in the whole combined data with the same threshold, we convolved all frames with two-dimensional Gaussian with appropriate $\sigma$ so that FWHMs of PSF at any place in the mosaicked image become the same (namely, the largest value of $\sim$ 6.0 arcseconds). 
The detection threshold is set to 25.5 mag\,arcsec$^{-2}$ in the $B$ band. 
This corresponds to 1.5-3 $\sigma _{\rm sky}$ above the sky level. 
If more than 10 pixels at which the counts exceed the threshold are connected, we regard them as an object. 
Altogether 6822 objects were detected. 
They consist of stars, galaxies, sky noises, and junks. 
All the above procedures are performed almost automatically by the data reduction software system developed by our group (Doi et al. 1995). 

The error in astrometry is 0.9 arcsecond in rms (with 2$\sigma$ rejection), the magnitude zero-point error is 0.02 mag. in rms, and the random error in magnitude is 0.2 mag. in rms (Akiyama 1996). 
The large random error in magnitude is due to bad weather conditions, namely large fluctuation of seeing size. 
However, as described in the last section, these errors are within tolerance for our cluster-finding technique. 

\subsubsection{Star/galaxy discrimination}

The detected objects consist of stars, galaxies, sky noises, and junks. 
We extract galaxies from the objects using the photometrical information, namely `sharpness' of the image and magnitude. 

Fig. 8 shows the distribution of all objects in the `sharpness'-magnitude diagram. 
`Sharpness' is defined by $I_{\rm peak}/\sqrt{N_{\rm pix}}$, where $I_{\rm peak}$ means the peak count of an object and $N_{\rm pix}$ means the number of pixels belonging to the object. 
In Fig. 8, we can recognize a tight sequence, which corresponds to stars. 
The bending of the sequence at the bright end reflects the saturation of CCDs. 
Galaxies, having flatter profile than stars with the same magnitude, are widely distributed in the region below the star sequence. 
Most of sky noises occupy faint and unsharp end of the diagram (often in several short sequences) and we can eliminate them by simply setting a cut-off magnitude. 
Other brighter unquestionable junks with quite flat profiles, due to bad pixel columns, haloes around bright stars, and sometimes loci of artificial satellites, often mingle galaxies. 
These junks must be carefully checked and removed. 

\begin{figure}
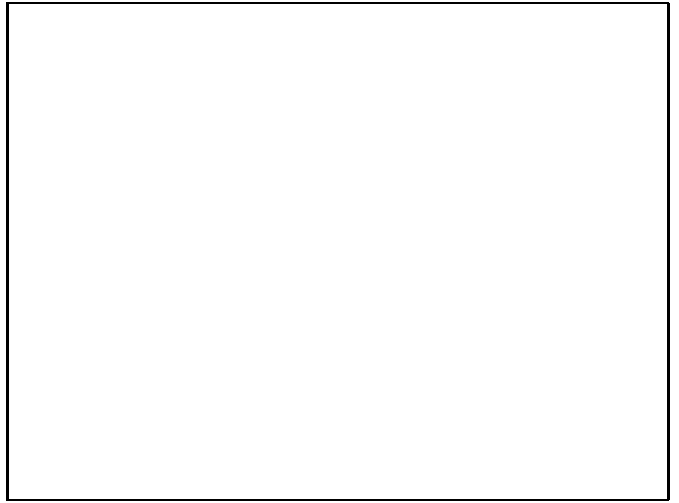

\picplace{6.6cm}
\caption[ ]{Star/galaxy discrimination diagram for the 6822 `object's in the $B$ band MCCD data in the NGP region. The border between `star region' and `galaxy region' is shown as a solid line.}
\end{figure}

We separate galaxies from the other objects with the following boundaries. 
The first one is a line corresponding to Gaussian profiles with $\sigma$=1.1$\sigma_{\rm PSF}$, where $\sigma_{\rm PSF}$ is the standard deviation of the best-fit Gaussian to the PSF that was composed of stellar images. 
PSFs have usually longer tails than that of Gaussian and are never well-fitted with a single Gaussian profile. 
However, for fainter magnitudes, outskirts of PSFs become negligible, and an approximation with single Gaussian is good enough. 
The second boundary is $I_{\rm peak}/\sqrt{N_{\rm pix}} = 300$. 
Actually, some objects above this boundary and below the star sequence are blended objects; in most cases, they are galaxies overlapping with stars. 
As magnitude goes fainter, the star sequence falls and eventually merges into the galaxy territory. 
And so do sky noises. 
It is no longer possible to discriminate between stars and galaxies. 
Therefore, our galaxy sample must also be restricted by the limit of star/galaxy discrimination. 
We fix the limit to be $m_{\rm limit}$ = 21.0, which is the third boundary. 

Finally, we cut off the uneven edge region due to the dead CCD chip and select the central 1.7$\times$2.9 deg$^2$ rectangular region which contains 996 galaxies. 
The two-dimensional distribution of these galaxies are shown in Fig. 9. 
North is up and east is to the left. 

\subsection{Results}

Fig. 10 shows the `richness image' of the NGP region for $z_{\rm fil}=0.20$. 
We can find some cluster candidates as peaks. 
Table 1 lists 18 significant peaks with $N_{\rm p} > 180$ at $z_{\rm fil} = 0.20$, and their ${\cal L}_{\rm p} - z_{\rm fil}$ curves are shown in Fig. 11. 
In Fig. 11, only No.3 and 7 have a prominent single peak in their ${\cal L}_{\rm p}-z_{\rm fil}$ curves. 
The ${\cal L}_{\rm p}-z_{\rm fil}$ curves for the other cluster candidates are almost flat and featureless (No.4, 5, 8, 13, and 18) or monotonically descends as the filter redshift increases (the others). 

\begin{figure}
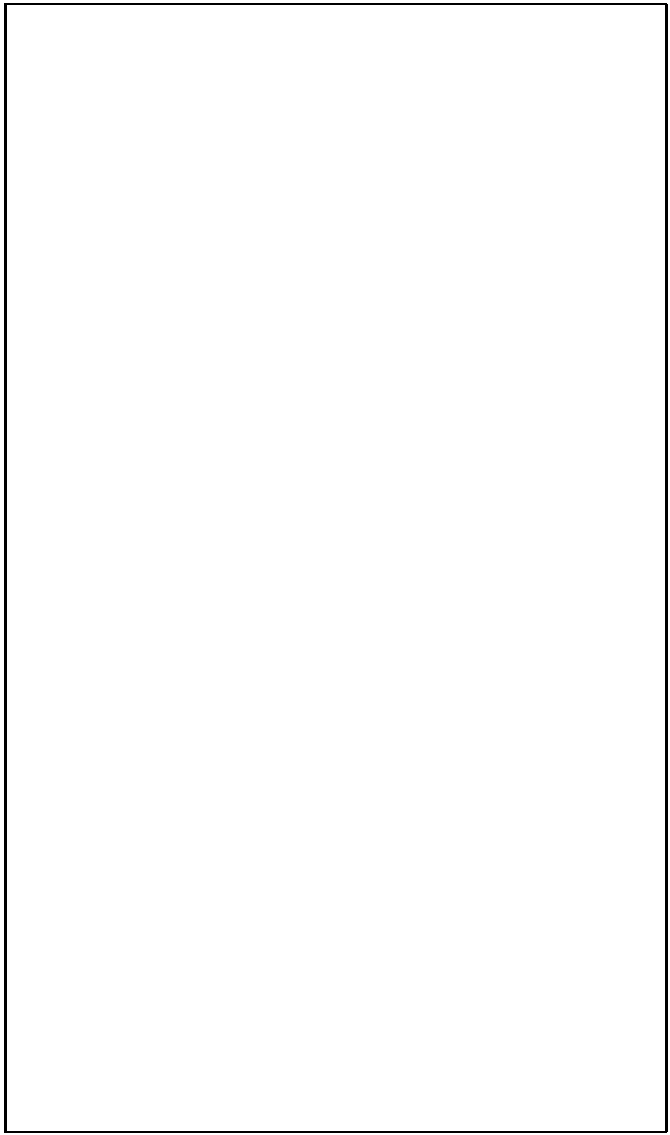

\picplace{15cm}
\caption[ ]{The distribution of 996 galaxies in the 1.7$\times$2.9 deg$^2$ region. Symbol size changes with apparent magnitude. The largest symbol corresponds to $m_B$ = 14.2 and the smallest to $m_B$ = 21.0. The field center is at ($\alpha , \delta$) = ($13^{\rm h}08^{\rm m}55^{\rm s}.3, +30^{\circ}00'47''.2$) (J2000.0). North is up and east is to the left.}
\end{figure}

An ${\cal L}_{\rm p}-z_{\rm fil}$ curve which monotonically descends with increasing filter redshift does not always mean that the redshift of the corresponding cluster candidate is less than 0.1; in the most cases, ${\cal L}_{\rm p}-z_{\rm fil}$ curves just keep increasing and have no peak, or become noisy, as filter redshift becomes even smaller. 
These behaviors are similar for the case of monotonically ascending ${\cal L}_{\rm p}-z_{\rm fil}$ curve. 
Thus, most of the cluster candidates with flat or monotonically descending/ascending  ${\cal L}_{\rm p} - z_{\rm fil}$ curves are spurious. 

\begin{figure}
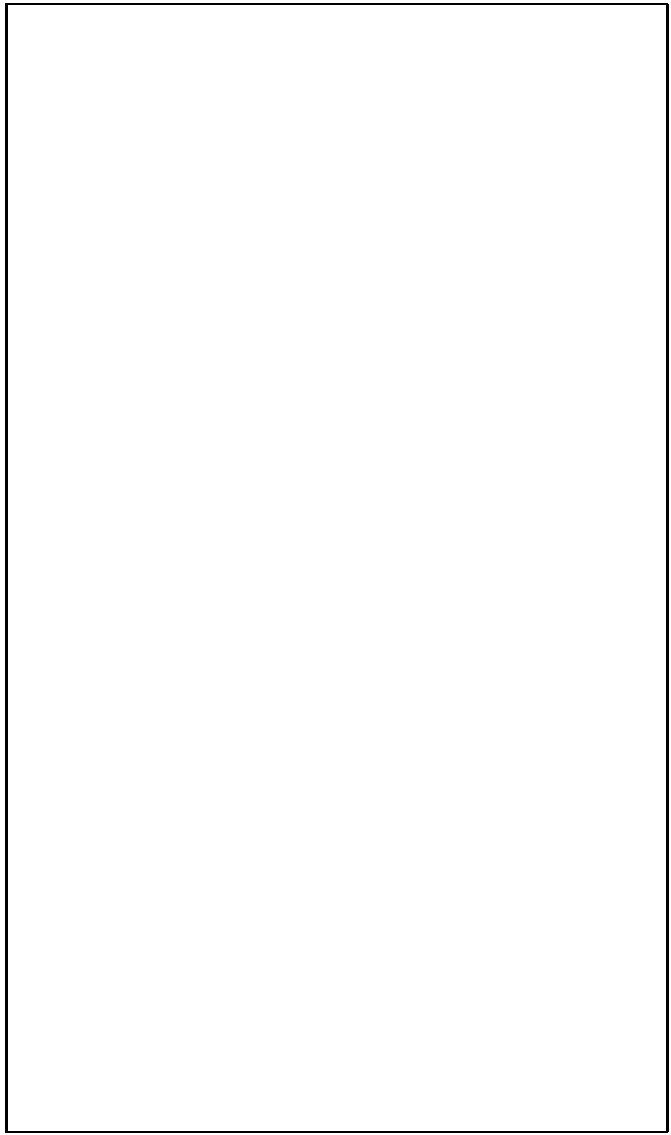

\picplace{15cm}
\caption[ ]{`Richness image' of the NGP galaxy sample of Fig. 9 for $z_{\rm fil}$ = 0.20. Solid line shows the contour of the threshold richness $N_{\rm th}$ = 180. Plus signs indicate positions of cluster candidates.}
\end{figure}

We can recognize many cluster candidates gathered in the bottom-right region in Fig. 10. 
However, the area includes the north-eastern outskirts of the Coma cluster, which correspond to the concentration of bright galaxies in the bottom-right region in Fig. 9. 
No cluster candidates in this region have a single peak in their ${\cal L}_{\rm p} - z_{\rm fil}$ curves, implying that most of them may be spurious. 
The ${\cal L}_{\rm p} - z_{\rm fil}$ curves of the candidates No.16-18 do not show a single peak and their $N_{\rm p}$ values are too small (less than 100). 
They may also be spurious. 

The most significant cluster candidates are No.3 and 7. 
These candidates both correspond to a single Abell cluster A1677. 
Splitting into two peaks may be either due to the poor quality (for example, the bright limiting magnitude or the inhomogeneity) of the data or due to a possible substructure. 
The measured redshift of A1677 is 0.183 (ACO) and the Abell richness $c$ is 112, which corresponds to $N\sim$ 3000. 
Our estimates of ($z$, $N$) are (0.26, 697.1) for No.3 and (0.16, 673.5) for No.7. 
There is another cataloged cluster, No.14. 
This cluster is II Zw 1305.4+2941 (Koo et al. 1986 and references therein). 
It has also been detected with X-ray satellites such as {\em Einstein} (MS 1305.4+2941 in Gioia et al. 1990), {\em ROSAT} (1RXS J130749.3+292536 in Voges et al. 1996), and {\em ASCA} (Ueda 1996). 
The measured redshift of this cluster is 0.241, while our redshift estimation for this cluster gives 0.10. 
Taking into account the poor quality of the data and the brighter limiting magnitude than that of the simulations in Sect. 3, we conclude that redshift and richness estimations for these two clusters are consistent with the cataloged values. 

Let us compare this result with that of intuitive eye selection. 
A glance of the galaxy distribution in Fig. 9 can find some other `somewhat conspicuous' galaxy concentrations. 
They are, for example, at (X,Y) = (35,110), (40,90), and (60,80). 
These three clumps seem to be more plausible `clusters' at a glance than the fainter one, for example, No.14 in Table 1 at (X,Y) = (55,55). 
However, when we examine a `richness image' (Fig. 10), these three appear to have much less remarkable peaks than No.14, which is a real cluster. 

Searching for clusters with a simultaneous use of magnitudes and positions can produce a quite different, and more objective result than that by conventional techniques using surface density of galaxies only. 

\section{Conclusion}

We have developed an objective and automatic cluster-finding method. 
It is a variant of P96 method with some improvements. 
The method uses positions and apparent magnitudes of galaxies simultaneously, and detects clusters by fitting artificial cluster models which contain redshift and richness as free parameters by maximizing the likelihood function. 
Therefore redshift and richness of clusters are estimated as byproducts of detection. 
Good accuracies in the estimates of cluster's position, redshift, and richness are confirmed by a number of Monte Carlo simulations. 
For clusters at $z$=0.20 and as rich as the Coma cluster, errors in estimating redshift and richness are $\Delta z\sim$0.02 and $\Delta N\sim$360 (12\%), respectively. 
Spurious detection rate of this method is also studied with Monte Carlo simulations and is shown to be less than $\sim$10\% of that by conventional techniques using only surface density of galaxies. 
A cluster survey in the NGP region is performed as a test with real data. 
Despite the poor quality of the data, two known real clusters are successfully detected. 

\begin{table}
\caption[ ]{All detected cluster candidates in the NGP region.}
\begin{flushleft}
\begin{tabular}{llllll}
\hline
\hline
No. & $\alpha$(2000.0) & $\delta$(2000.0) & $z_{\rm est}$ & $N_{\rm est}$ & Notes.\\
\hline
 1 & 13 05 39.1 & 30 51 36.5 & 0.10    & 157.7    & $\dagger$ \\
 2 & 13 05 43.3 & 28 58 33.0 & $<$0.1? & $<$127.8 & $\dagger$ \\
 3 & 13 05 47.5 & 31 14 38.1 & 0.26    & 697.1    & A1677 \\
 4 & 13 05 48.7 & 28 47 33.5 & $<$0.1? & $<$84.7  & $\dagger$ \\
 5 & 13 05 49.2 & 28 37 03.5 & $<$0.1? & $<$386.1 & $\dagger$ \\
 6 & 13 05 59.4 & 29 08 04.8 & 0.11    & 175.9    & $\dagger$ \\
 7 & 13 06 03.6 & 31 20 40.0 & 0.16    & 673.5    & A1677 \\
 8 & 13 06 07.2 & 29 45 36.7 & $<$0.1? & $<$69.2  & $\dagger$ \\
 9 & 13 06 13.5 & 28 56 36.0 & $<$0.1? & $<$136.9 & $\dagger$ \\
10 & 13 06 25.1 & 28 51 37.0 & $<$0.1? & $<$156.8 & $\dagger$ \\
11 & 13 06 52.0 & 29 10 09.5 & $<$0.1? & $<$100.8 & $\dagger$ \\
12 & 13 07 06.0 & 28 58 40.3 & 0.11    & 124.5    & $\dagger$ \\
13 & 13 07 17.3 & 29 07 11.2 & 0.14    & 116.0    & $\dagger$ \\
14 & 13 08 25.7 & 29 26 45.0 & 0.10    & 94.6     & $\ast$ \\
15 & 13 09 18.7 & 28 40 45.2 & $<$0.1? & $<$120.8 & $\dagger$ \\
16 & 13 10 02.5 & 30 55 19.7 & $<$0.1? & $<$100.3 & $\ddagger$ \\
17 & 13 10 07.2 & 31 06 20.0 & $<$0.1? & $<$88.6  & $\ddagger$ \\
18 & 13 10 55.9 & 30 37 47.6 & $<$0.1? & $<$78.6  & $\ddagger$ \\
\hline
\end{tabular}
\end{flushleft}
$\dagger$ -- may be junk (edge of Coma)\\
$\ddagger$ -- may be junk\\
$\ast$ -- II Zw 1305.4+2941
\end{table}

\begin{figure}
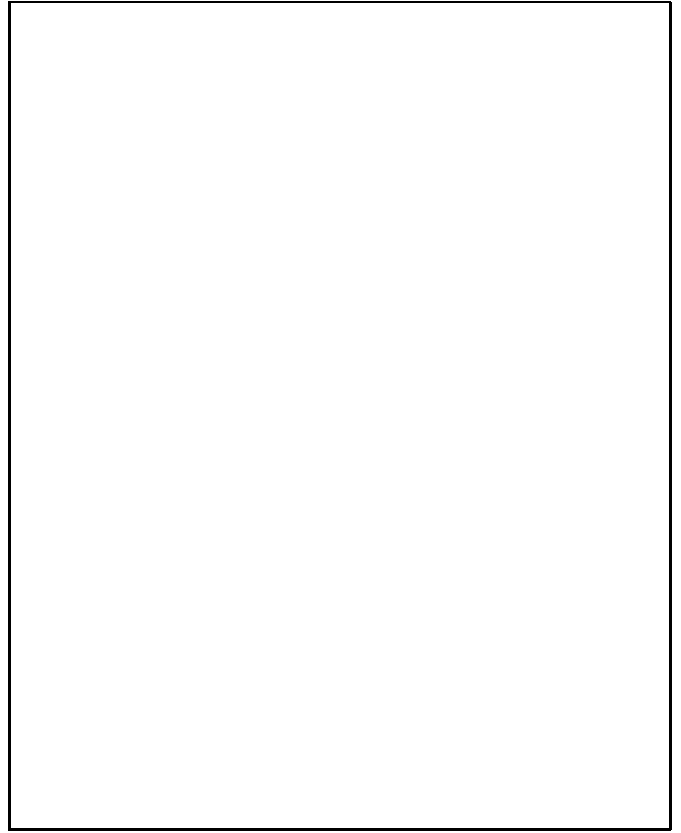

\picplace{11cm}
\caption[ ]{Peak logarithmic likelihood as a function of filter redshift for the 18 cluster candidates shown in Fig. 10.}
\end{figure}

At present, it is quite difficult to make a deep X-ray survey over a wide area on the sky and to build a large catalog of, especially, distant ($z>0.3$) clusters though some attempts are being made such as SHARC (Collins et al. 1997; Burke et al. 1997), RDCS (Rosati et al. 1998), RIXOS (Castander et al. 1995), and WARPS (Scharf et al. 1997). 
Thus optical search is almost the only realistic way to find a large number of distant clusters. 
Objective and automated methods for finding clusters from optical data must be indispensable tools in the near future for quickly constructing large, statistically complete cluster catalogs from the data covering an extremely wide area, for example, those from SDSS (Gunn and Weinberg 1995; Okamura 1995), or for compiling catalogs of extremely distant clusters up to $z>1$ from deep imaging data with 8-10m class telescopes. 

\begin{acknowledgements}

The authors would like to thank Naoki Yasuda and Masafumi Yagi for valuable comments and suggestions in the early stage of this work and Maki Sekiguchi for the Mosaic CCD camera development and help in observation. 
WK is grateful to Nobunari Kashikawa, Toru Yamada, and Masayuki Akiyama for their help in reducing the NGP data. 

\end{acknowledgements}

\appendix

\section{Relationship between $N$ and Abell richness}

We adopt $N$, the field-corrected number of all galaxies brighter than $m^*$+5, as the indicator of cluster richness in this paper. 
In most previous studies, however, Abell's richness parameter $c$, which is the field-corrected number of galaxies inside the Abell radius with magnitude between $m_3$ and $m_3+2$, is conventionally used. 
For convenience, we estimate the relationship between $N$ and $c$ by simply using random values whose probability density function obeys a Schechter function. 

For a given value of $N$, we generate 20 `clusters' and count numbers of galaxies with magnitudes between $m_3$ and $m_3+2$. 

Fig. 12 shows the relationship. 
The median of 20 $c$ values corresponding to a given $N$ is shown by solid line. 
The region between the two dashed lines represents the central $\pm 1\sigma$ area. 
The median value of $c$ is well approximated by the following power-law relations 
\begin{eqnarray}
c & = & 0.46\; N^{0.67} \hspace*{0.5cm} (N\la 4000) \\
c & = & 0.33\; N^{0.71} \hspace*{0.5cm} (N\ga 4000)
\end{eqnarray}
For the Coma cluster $c$ equals 106, which corresponds to $N\sim$ 3000. 

The magnitude of the third brightest galaxy is often significantly affected by the overlapping galaxies or variations in cluster morphology; if a cluster's morphology is cD or B in Rood-Sastry classification (Rood and Sastry 1971), $m_3$ would be often fainter because the original third brightest galaxy may have been merged by the first or second brightest galaxy and a fainter galaxy becomes the new third brightest galaxy. 
On the other hand, $m_3$ would be brighter for L or C clusters because the third brightest galaxy may capture smaller ones. 
The error in estimating $m_3$ directly affects $c$ and leads eventually to some serious systematic error in the evaluation of cluster richness. 
On the other hand, $N$ includes {\em all} galaxies brighter than $m^*$+5 and is less affected by those factors. 
We can use $N$ as a more robust richness indicator than $c$. 

\begin{figure}
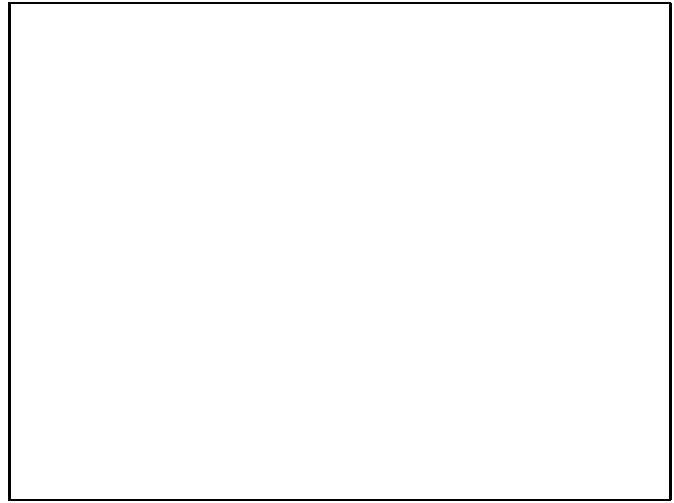

\picplace{6.6cm}
\caption[ ]{Relationship between our richness parameter $N$ and Abell richness $c$. Solid line shows the median value of $c$ of 20 artificial clusters with a fixed $N$. Dashed lines show $\pm 1\sigma$ range.}
\end{figure}

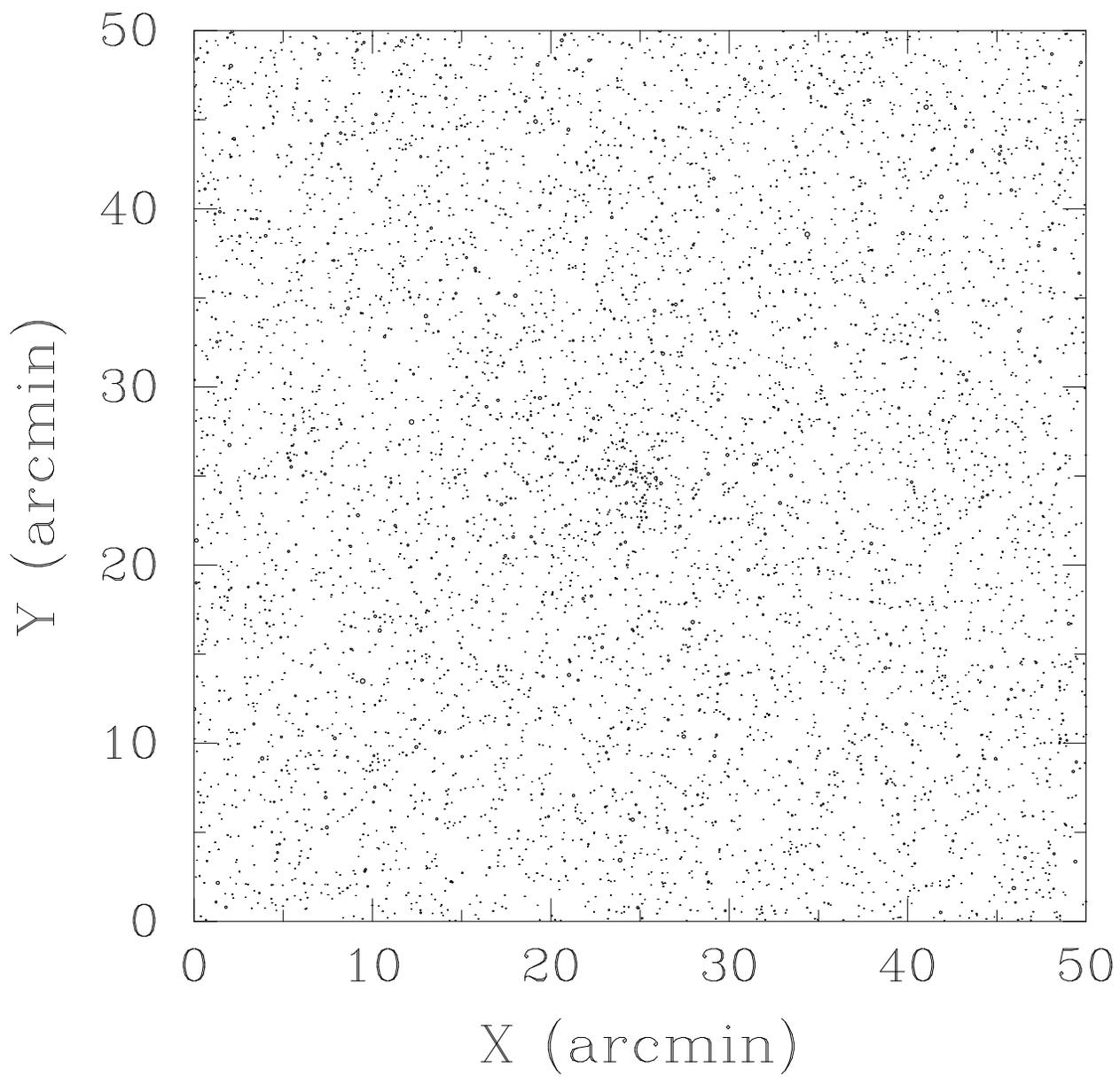
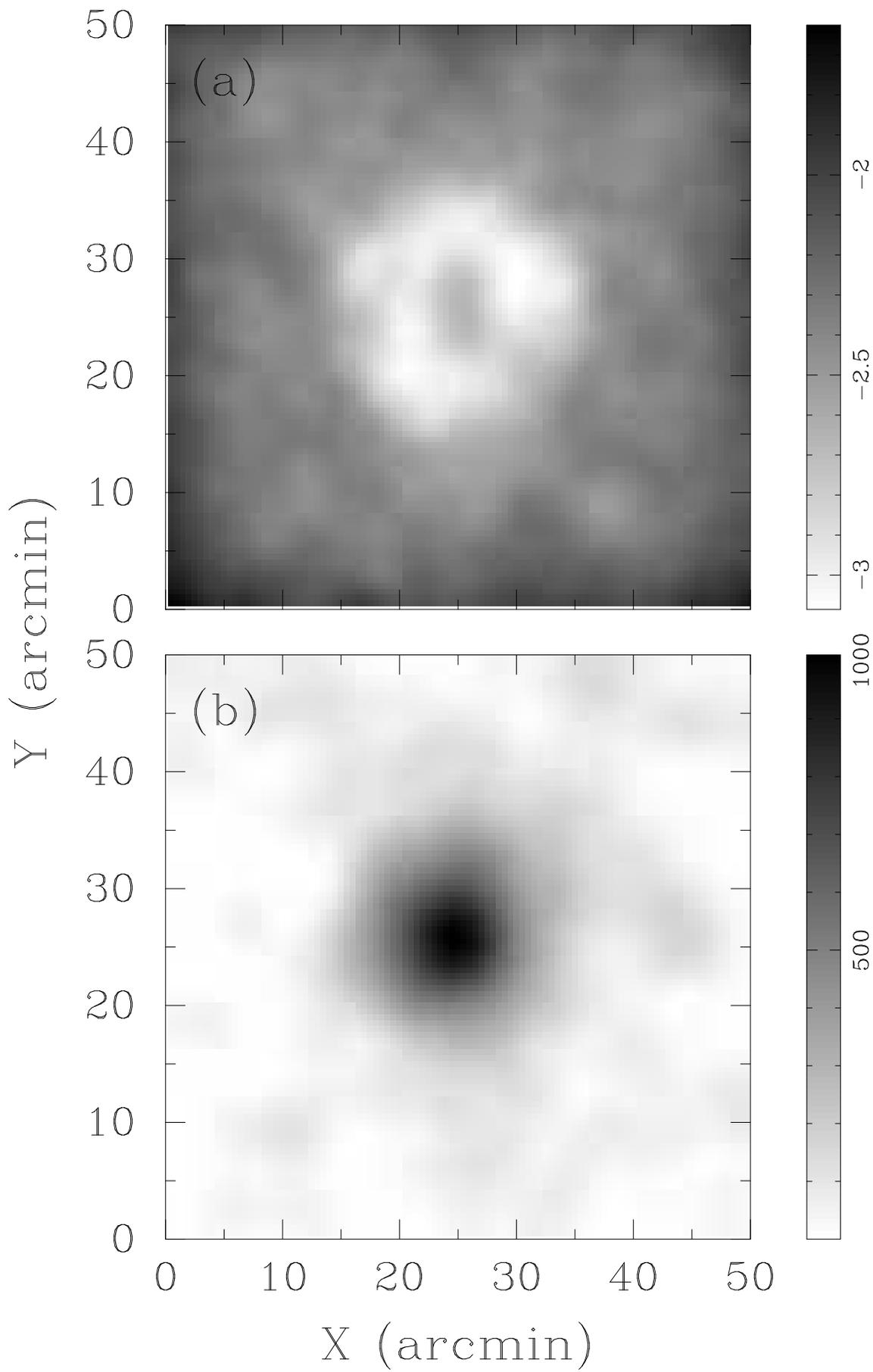
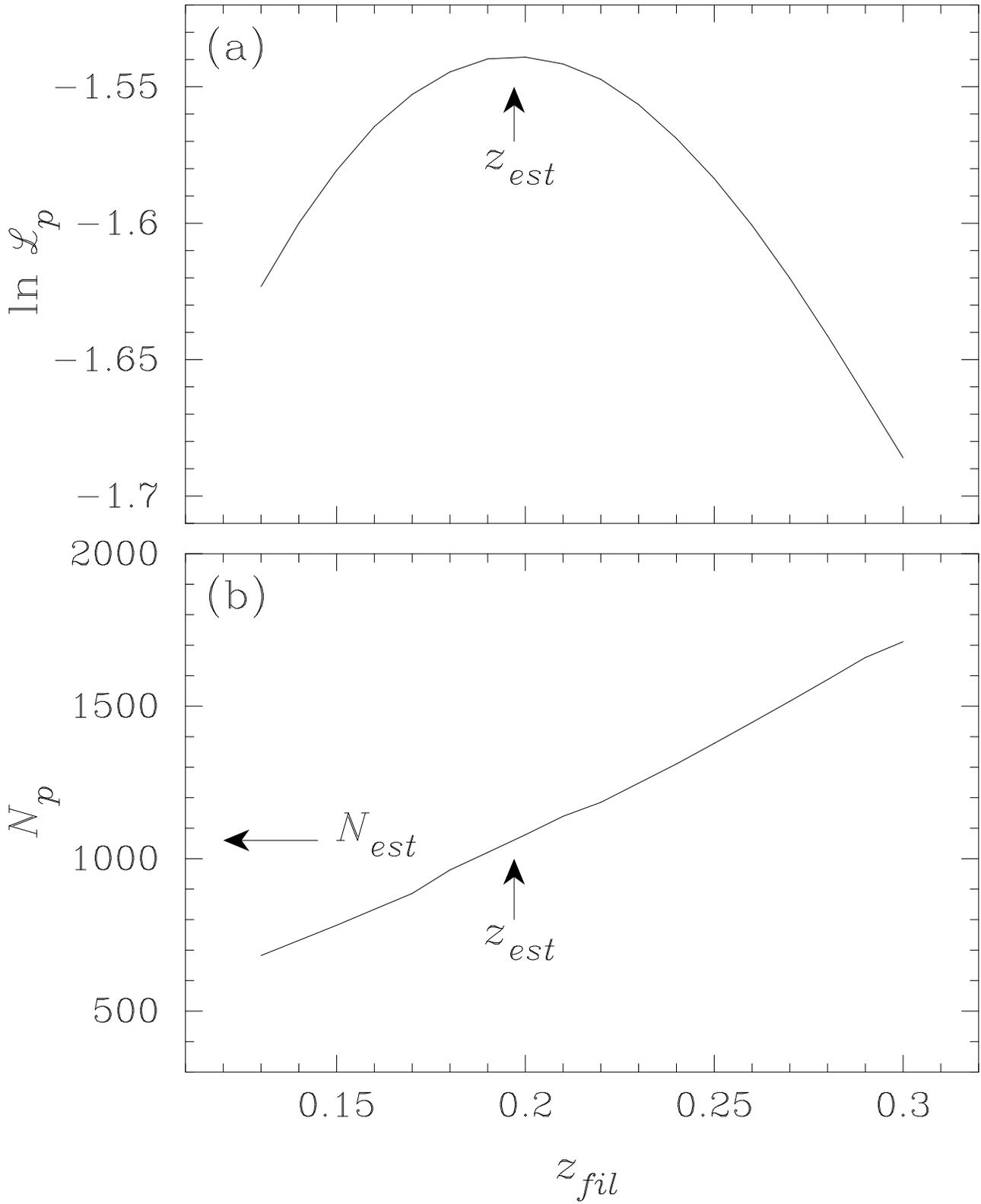
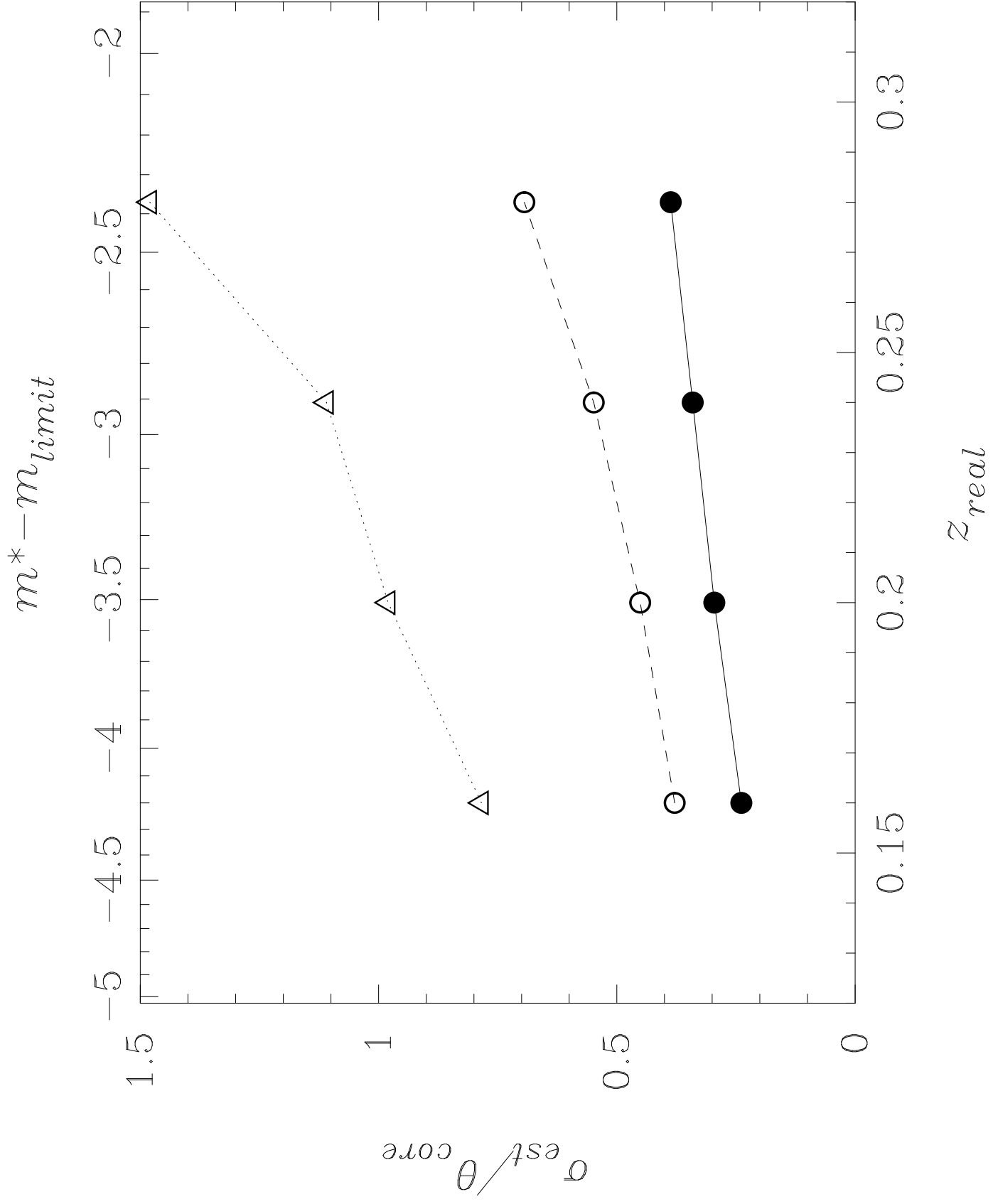
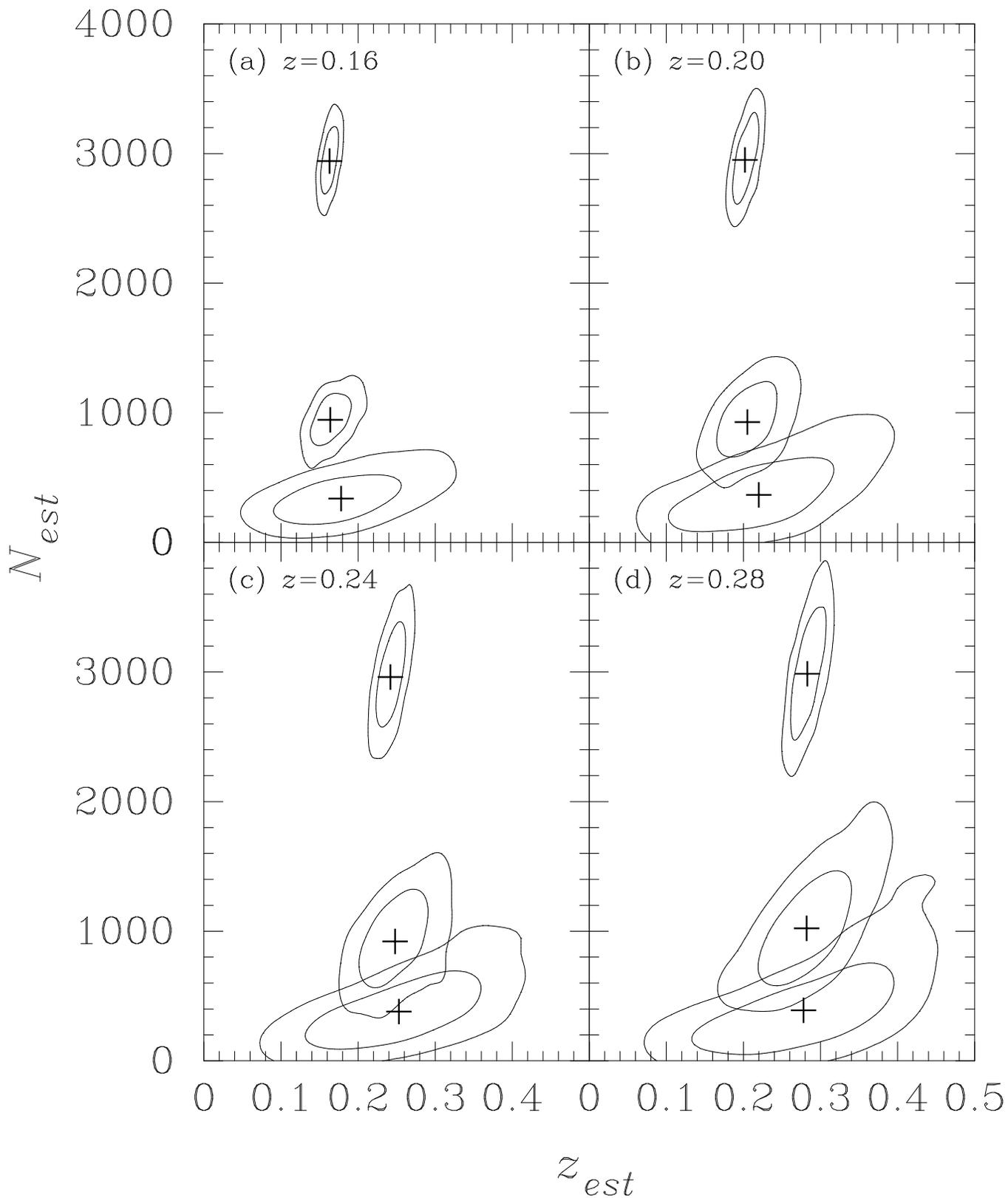
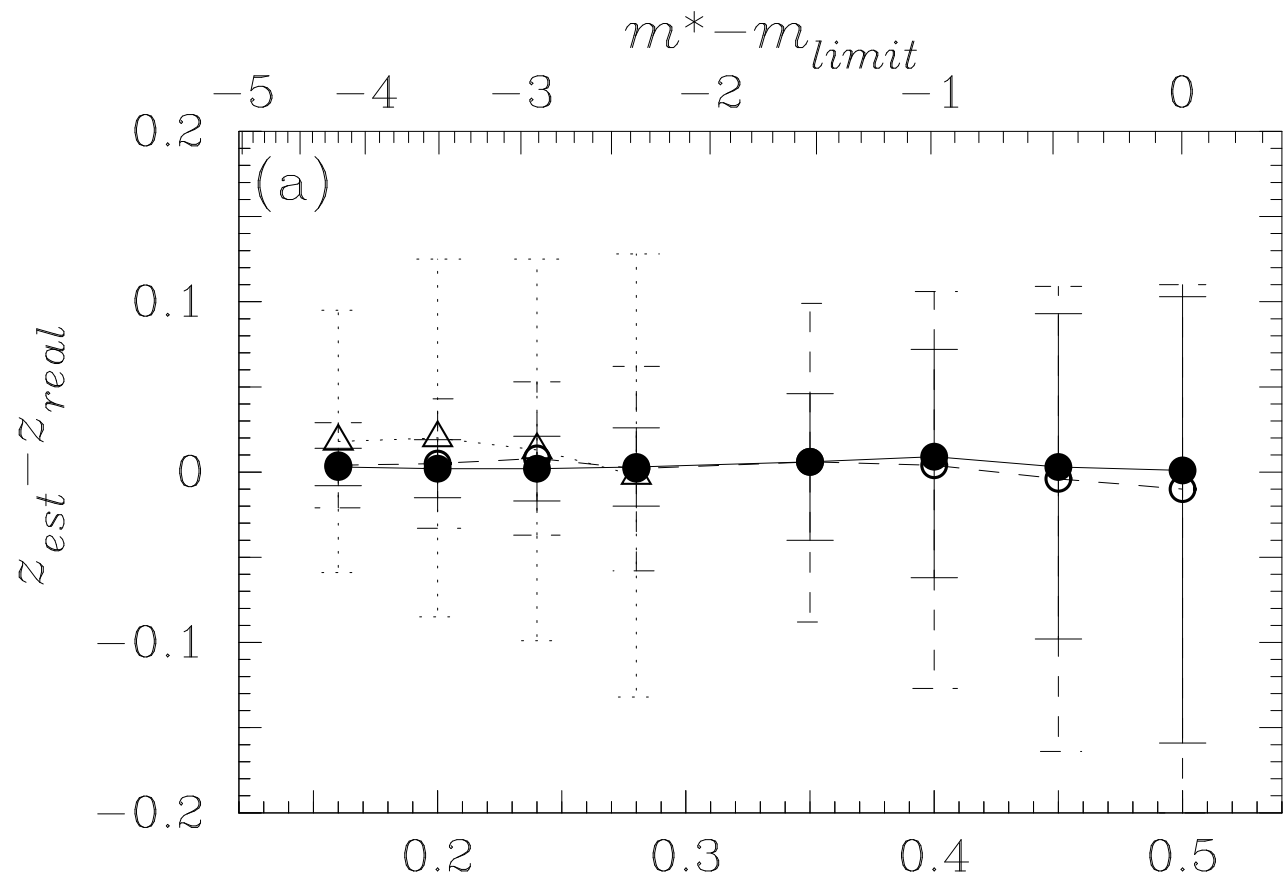
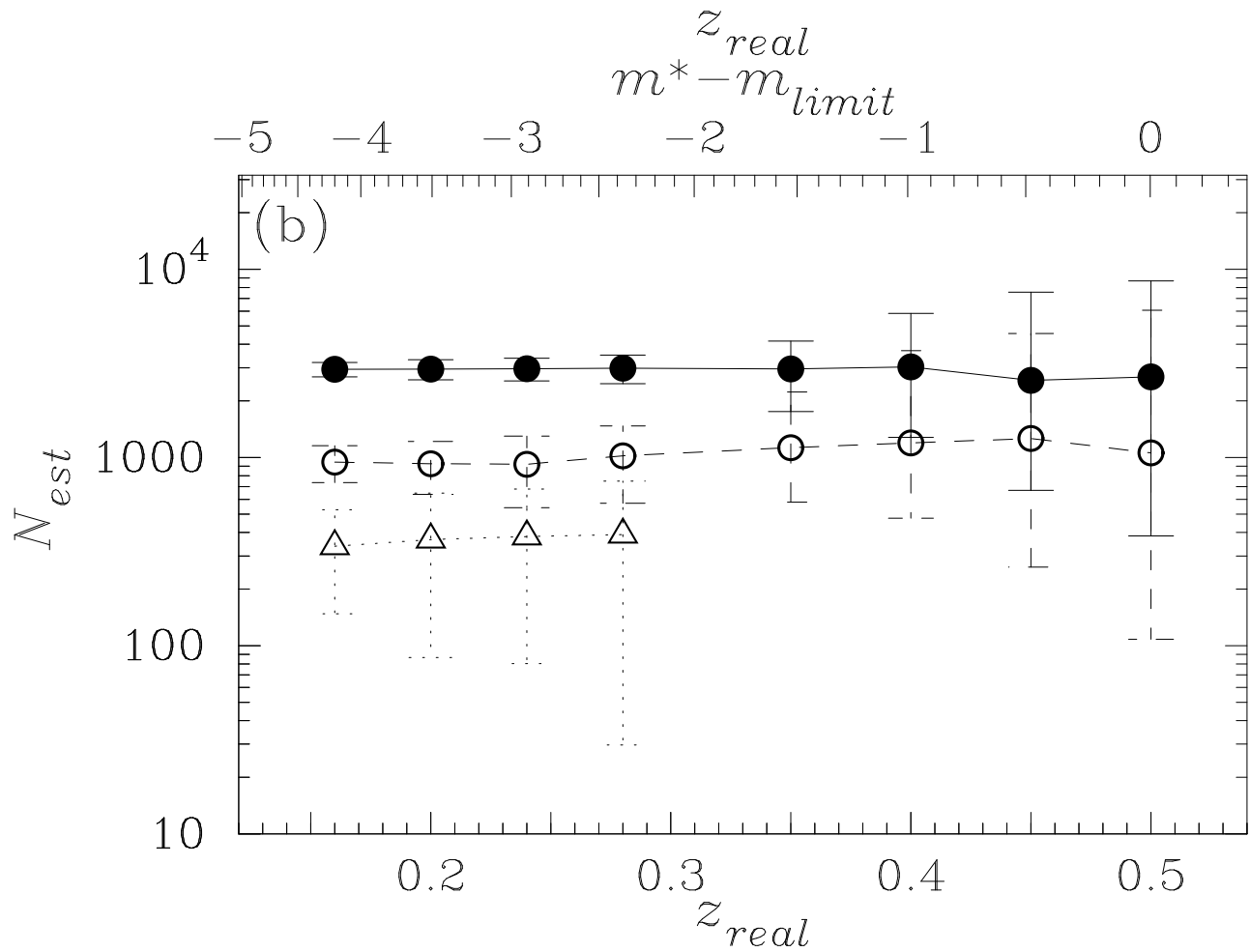
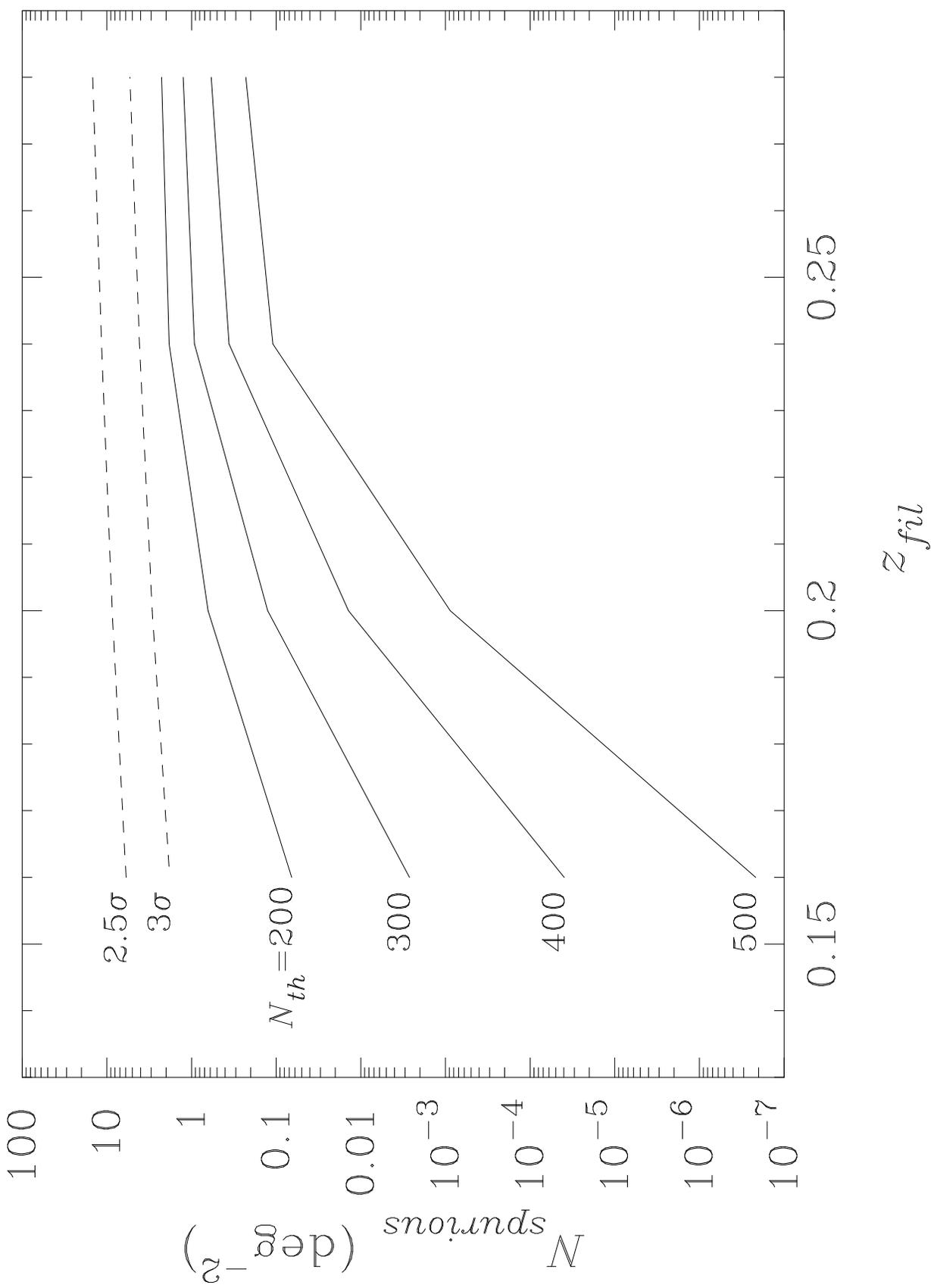
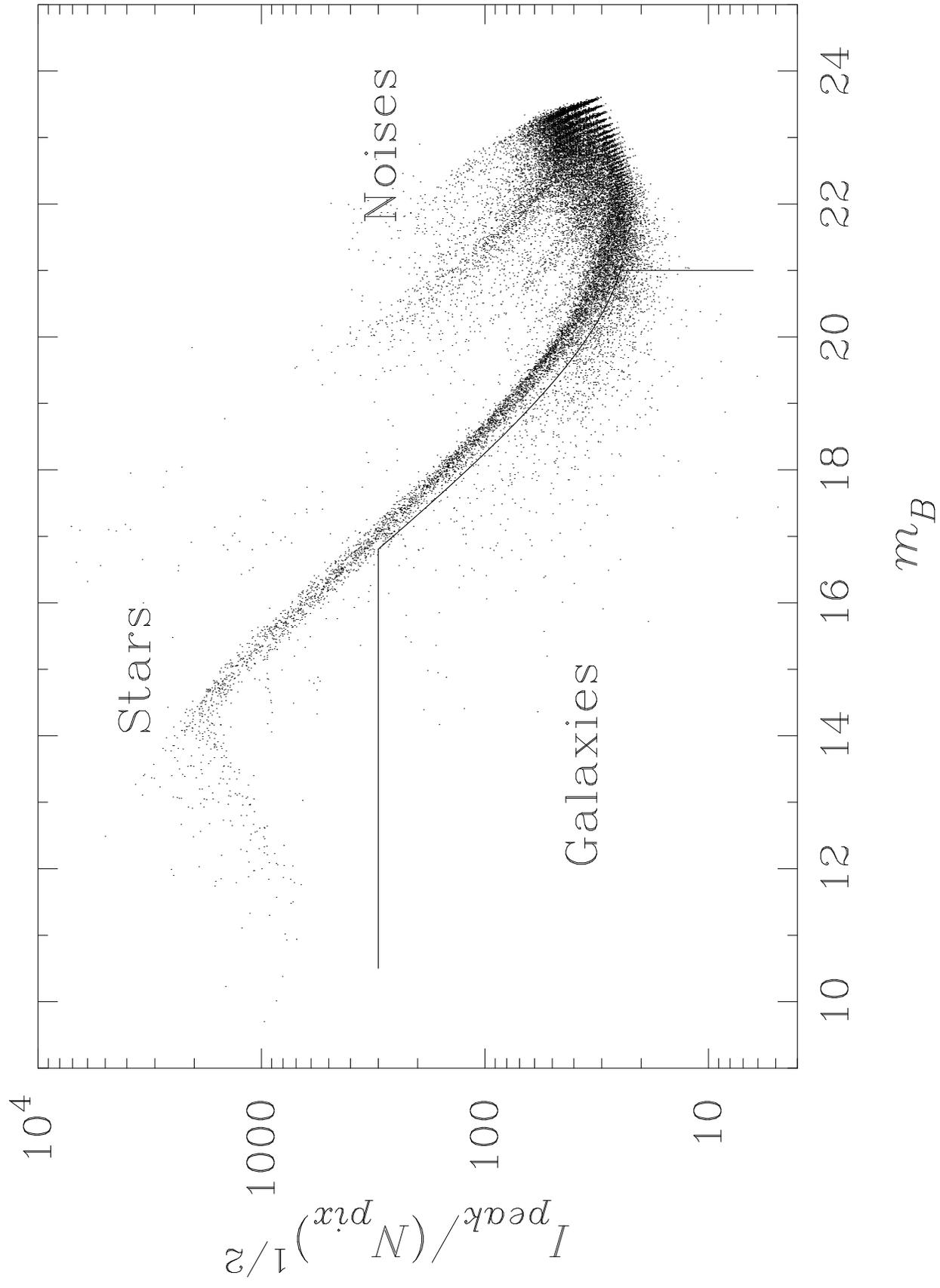
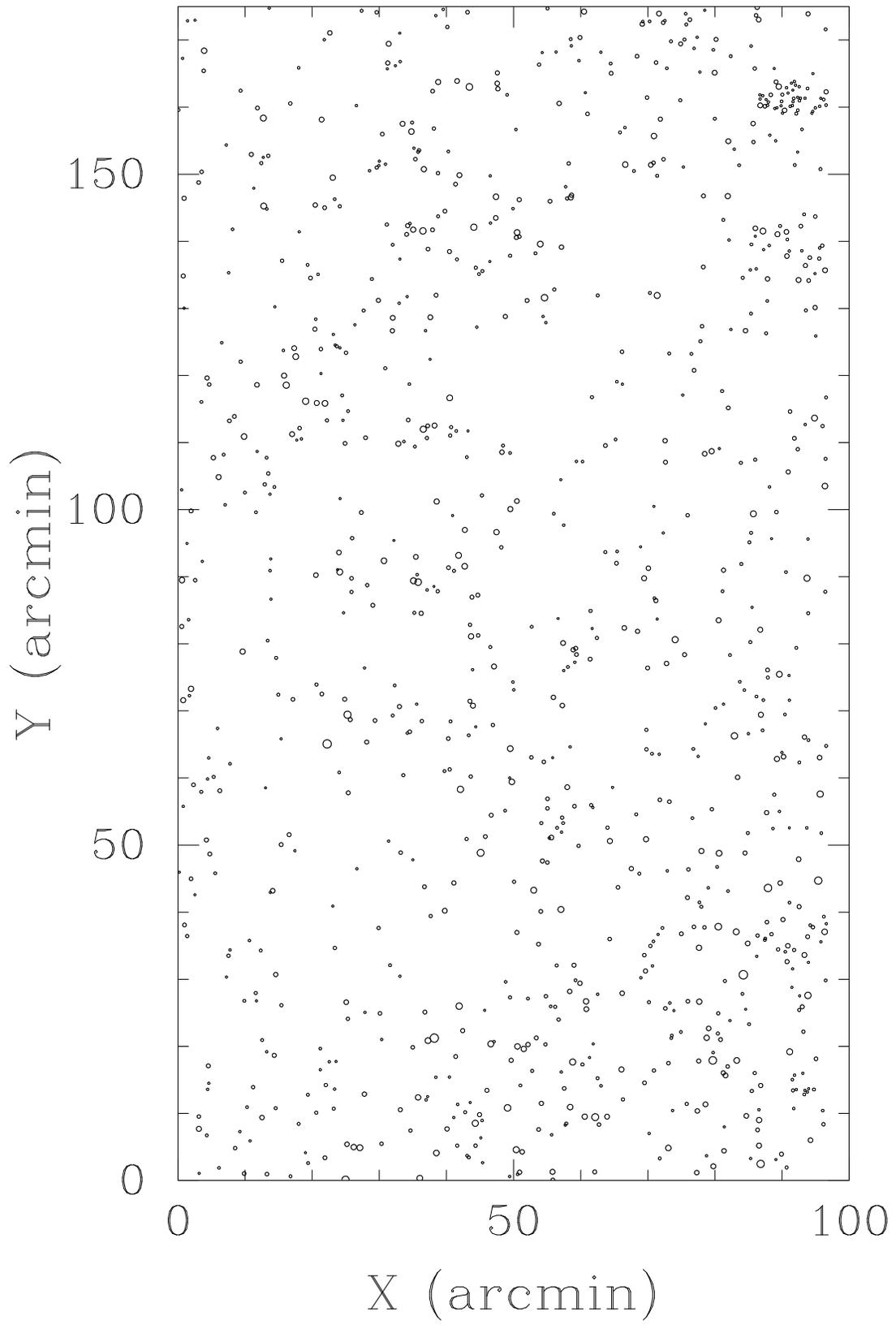
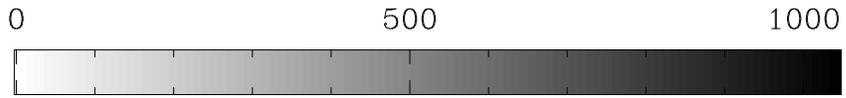
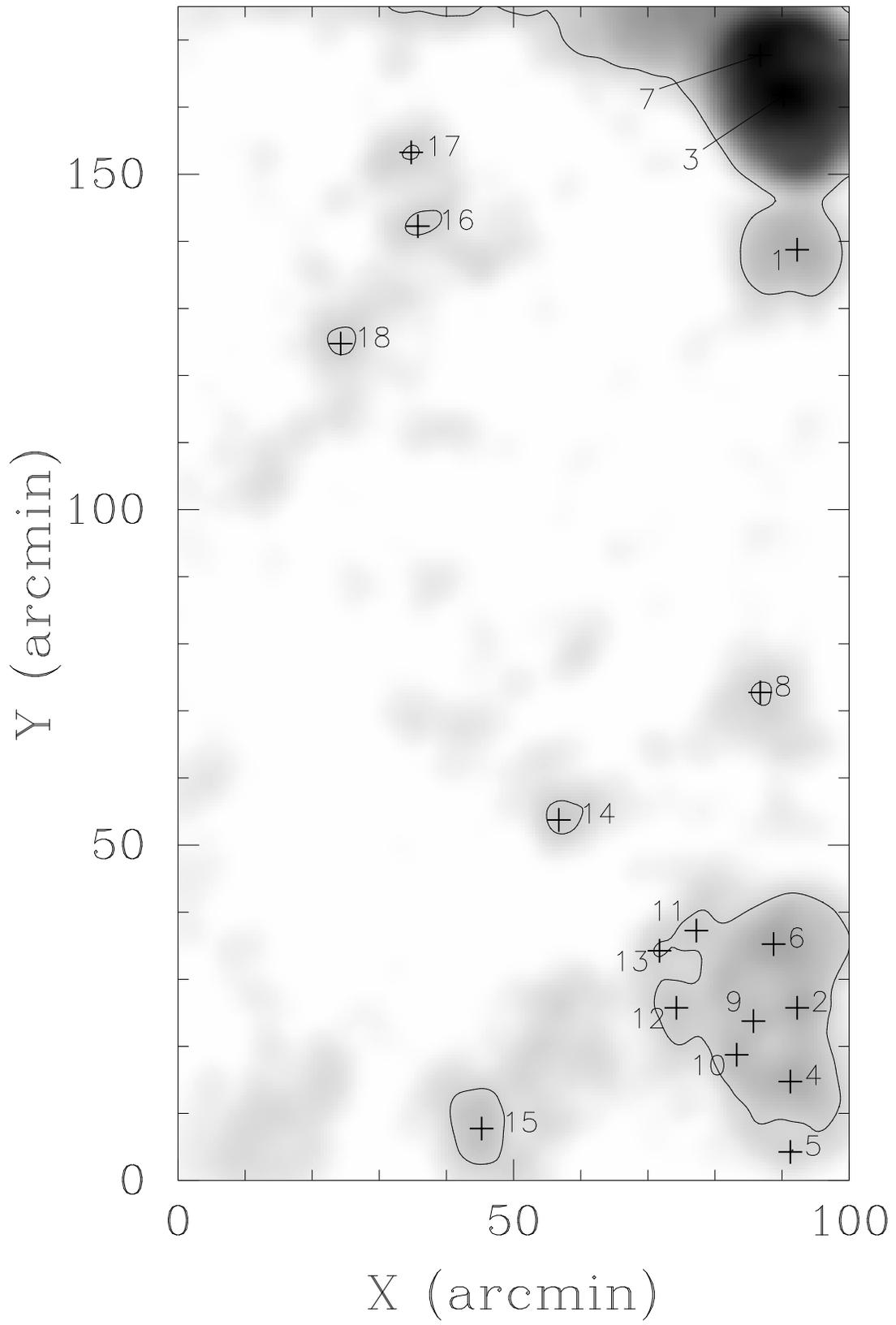
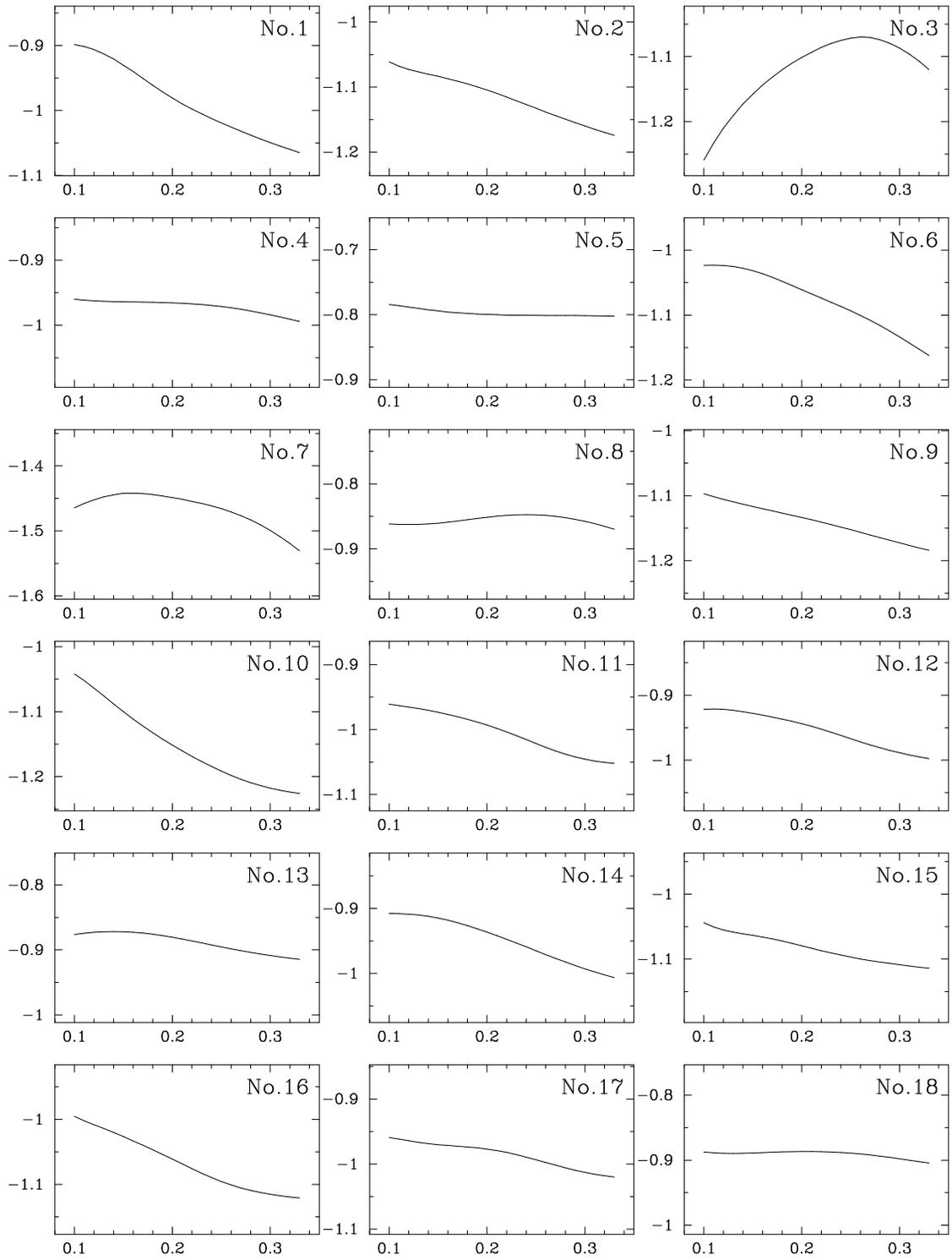
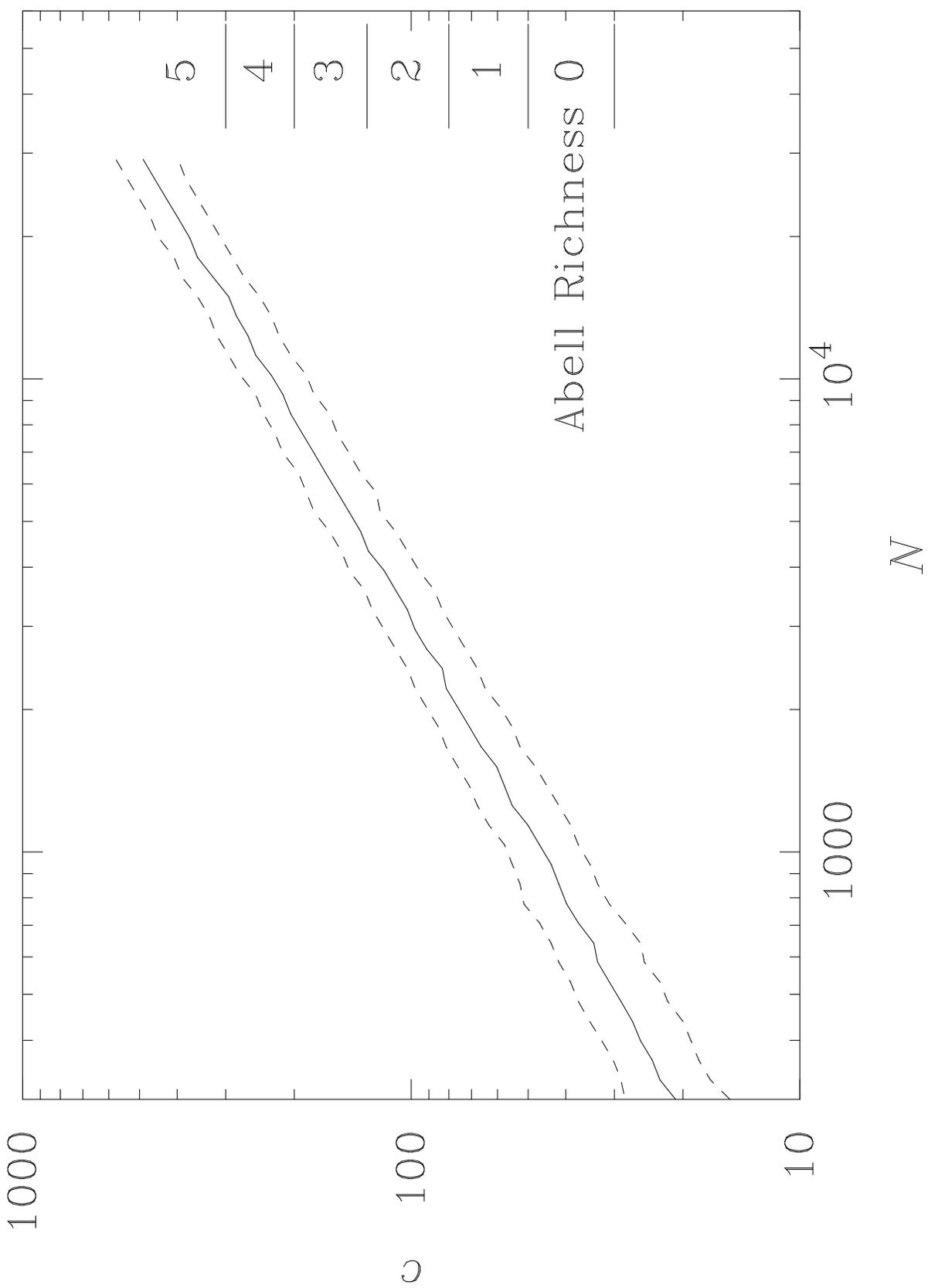
\end{document}